\PassOptionsToPackage{dvipsnames}{xcolor}
\documentclass[twocolumn]{aastex631}
\usepackage{amsfonts,amsmath,appendix,graphics,graphicx,color,hyperref}
\usepackage{gensymb}

\shorttitle{Late-time HST/JWST Observations of GRB\,221009A}
\shortauthors{Sears et al.}

\usepackage[dvipsnames]{xcolor}
\begin{document}


\author[0000-0001-8023-4912]{Huei Sears}
\affiliation{Department of Physics and Astronomy, Rutgers, the State University of New Jersey, 136 Frelinghuysen Road, Piscataway, NJ 08854-8019, USA}

\author[0000-0002-7706-5668]{Ryan Chornock}
\affiliation{Department of Astronomy, University of California, Berkeley, CA 94720-3411, USA}

\author[0000-0003-0526-2248]{Peter K. Blanchard}
\affiliation{Center for Astrophysics \textbar\ Harvard \&\ Smithsonian, 60 Garden Street, Cambridge, MA 02138-1516, USA}
\affiliation{The NSF AI Institute for Artificial Intelligence and Fundamental Interactions}

\author[0000-0003-4768-7586]{Raffaella Margutti}
\affiliation{Department of Astronomy, University of California, Berkeley, CA 94720-3411, USA}
\affiliation{Department of Physics, University of California, 366 Physics North MC 7300, Berkeley, CA 94720, USA}

\author[0000-0002-5814-4061]{V.~Ashley Villar}
\affiliation{Center for Astrophysics \textbar\ Harvard \&\ Smithsonian, 60 Garden Street, Cambridge, MA 02138-1516, USA}
\affiliation{The NSF AI Institute for Artificial Intelligence and Fundamental Interactions}

\author[0000-0002-2361-7201]{Justin Pierel}
\altaffiliation{NHFP Einstein Fellow}
\affiliation{Space Telescope Science Institute, Baltimore, MD 21218, USA}

\author[0000-0001-5661-7155]{Patrick J. Vallely}
\affiliation{Core Pricing Systems, CarMax, Inc., Richmond, VA 23238, USA}

\author[0000-0002-8297-2473]{Kate D.~Alexander}
\affiliation{Steward Observatory, University of Arizona, 933 North Cherry Avenue, Tucson, AZ 85721-0065, USA}

\author[0000-0002-9392-9681]{Edo Berger}
\affiliation{Center for Astrophysics \textbar\ Harvard \&\ Smithsonian, 60 Garden Street, Cambridge, MA 02138-1516, USA}
\affiliation{The NSF AI Institute for Artificial Intelligence and Fundamental Interactions}

\author[0000-0003-0307-9984]{Tarraneh~Eftekhari}
\altaffiliation{NHFP Einstein Fellow}
\affiliation{Center for Interdisciplinary Exploration and Research in Astronomy (CIERA), Northwestern University, 1800 Sherman Avenue, Evanston, IL 60201, USA}

\author[0000-0002-3934-2644]{Wynn ~V.~Jacobson-Gal\'{a}n}
\altaffiliation{NHFP Hubble Fellow}
\affil{Department of Astronomy and Astrophysics, California Institute of Technology, Pasadena, CA 91125, USA}

\author[0000-0003-1792-2338]{Tanmoy Laskar}
\affiliation{Department of Physics \& Astronomy, University of Utah, Salt Lake City, UT 84112, USA}
\affiliation{Department of Astrophysics/IMAPP, Radboud University, P.O. Box 9010, 6500 GL, Nijmegen, The Netherlands}

\author[0000-0002-2249-0595]{Natalie LeBaron}
\affiliation{Department of Astronomy, University of California, Berkeley, CA 94720-3411, USA}

\author[0000-0002-4670-7509]{Brian D.~Metzger}
\affil{Department of Physics and Columbia Astrophysics Laboratory, Columbia University, New York, NY 10027, USA}
\affil{Center for Computational Astrophysics, Flatiron Institute, 162 5th Ave, New York, NY 10010, USA}

\author[0000-0002-0763-3885]{Dan Milisavljevic}
\affiliation{Department of Physics and Astronomy, Purdue University, 525 Northwestern Ave, West Lafayette, IN 47907, USA}


\title{Late-time HST and JWST Observations of GRB\,221009A: Evidence for a Break in the Light Curve at 50 Days}

\begin{abstract}
GRB\,221009A is one of the brightest transients ever observed with the highest peak gamma-ray flux for a gamma-ray burst (GRB).  A type Ic-BL supernova (SN), SN\,2022xiw,  was definitively detected in late-time JWST spectroscopy ($t = 195$ days, observer-frame). However, photometric studies have found SN\,2022xiw to be less luminous ($10-70\%$) than the canonical GRB-SN, SN\,1998bw. We present late-time Hubble Space Telescope (HST)/WFC3 and JWST/NIRCam imaging of the afterglow and host galaxy of GRB\,221009A at t  $\sim$ 185, 277, and 345 days post-trigger.  Our joint archival ground, HST, and JWST light curve fits show strong support for a break in the light curve decay slope at $t = 50 \pm 10$ days (observer-frame) and a supernova at $< 1.5 \times$ the optical/NIR flux of SN\,1998bw.  This break is consistent with an interpretation as a jet break when requiring slow-cooling electrons in a wind medium with the electron energy spectral index, $p >2$, and $\nu_m < \nu_c$.  Our light curve and joint HST/JWST spectral energy distribution (SED) also show evidence for the late-time emergence of a bluer component in addition to the fading afterglow and supernova. We find consistency with the interpretations that this source is either a young, massive, low-metallicity star cluster or a scattered light echo of the afterglow with a SED shape of $f_{\nu} \propto \nu^{2.0\pm1.0}$.
\end{abstract}

\section{Introduction}
Gamma-ray bursts (GRBs) are some of the most energetic and luminous events in the Universe.  The brightest GRBs usually have gamma-ray ($\sim 1$--$10,000$ keV) fluences on the order of $10^{-3}$ erg cm$^{-2}$ and peak fluxes on the order of $10^{-4}$ erg s$^{-1}$ cm$^{-2}$ \citep[see Tables 2 and 3 of][]{Burns2023}.  GRBs are also measured to have typical gamma-ray isotropic energies ($E_{\gamma, iso}$) on the order of $10^{53}$ erg \citep{Ajello2019, Frail2001} and textcolor{red}{isotropic} luminosities on the order of $10^{53}$ erg s$^{-1}$ \citep{Nava2012, Abbott2017, Tsvetkova2017, Burns2023}.

GRB\,221009A was discovered on 2022 October 09, 13:16:59.99 UT by the Fermi Gamma-ray Burst Monitor \citep{Veres22, Lesage23, Burns2023} and soon after by the Swift Burst Alert Telescope \citep[BAT, ][]{Dichiara2022, Williams2023}.  This GRB was detected to have a gamma-ray fluence of 0.19 erg cm$^{-2}$, peak flux of 0.01 erg s$^{-1}$ cm$^{-2}$, gamma-ray isotropic energy of $E_{\gamma, iso} = 10^{55}$ erg, and a peak isotropic-equivalent luminosity of $\sim 10^{54}$ erg s$^{-1}$ \citep{Burns2023, Lesage23}.  It is the record in gamma-ray fluence, $E_{\gamma, iso}$, and peak flux and is only superseded in isotropic-equivalent luminosity by two GRBs \citep{Burns2023}.  The GRB has low Galactic latitude, $b = 4^{\degree}.3$ \citep{Williams2023}, and suffers from high Galactic extinction \citep[$A_{V} \sim 4.2$ mag,][]{SF2011}. Despite this, the multi-wavelength afterglow was detected by a variety of observatories allowing for several extensive follow-up campaigns \citep{Huang22, Bright23, Fulton23, Levan2023, Laskar2023, LHASSO23, Negro2023, OConnor23, Shrestha23, Gokul2023, Blanchard2024, Kong2024, Rhodes2024}, and a redshift of $z=0.151$ was determined from VLT/X-Shooter spectroscopy of the afterglow \citep{deUgarte22a, Malesani23}.

Several previous studies presented synchrotron models of the afterglow \citep{Bright23, Fulton23, Guarini23, Laskar2023, Levan2023, Ren23, Sato23, Rhodes2024, Tak2024}.  In the synchrotron emission model, the GRB afterglow is initially expected to decay as a power-law in flux vs. time.  There is predicted to be at least one luminosity break in the light curve, called a ``jet break," which manifests as a steeper power-law decay in flux vs.\ time.  This break is due to relativistic beaming and the jetted nature of the source. The break time depends on the physical geometry (i.e., the jet opening angle, $\theta_{jet}$ , and the viewing angle); the intrinsic kinetic energy, $E_{k}$ (which can be parameterized to $E_{\gamma, iso}$ with assumptions about the energy efficiency of the GRB); the circumstellar density; and the density profile of the circumburst medium \citep{Rhoads1999, Sari1999, Chevalier2000, Frail2001, Bloom2003, Yi2015}.  Furthermore, the opening angle and the isotropic energy can be used to calculate the true gamma-ray energy release of the jet via $E_{\gamma} = E_{\gamma, iso} (1 -\cos(\theta_{jet}))$.  The superlatively high $E_{\gamma, iso}$ inferred from the gamma-ray prompt emission \citep{Burns2023, Frederiks2023, Lesage23, An2023} suggests a small jet opening angle (and therefore a small jet break time or dense circumburst medium) for realistic calculations of $E_{\gamma}$.  \cite{DAvanzo22, Shrestha23} reported evidence for a jet break at $t \sim $1 day (observer-frame).  \cite{Levan2023} presented observations which were interpreted to require a jet break at $t < 0.03$ days (observer-frame), in conflict with a jet break at $t\sim $1 day. \cite{Laskar2023} present a forward shock synchrotron model that can moderately explain the optical to X-ray data but cannot explain the radio and predicts a jet break at $t \sim 100$ days.   More data at later times and complete, multi-wavelength light-curve fitting is needed to resolve the seemingly conflicting results in many of these early works.

Collapsar-caused long GRBs are expected to be followed by a Type Ic-BL supernova \citep[SN,][]{WoosleyBloom2006}.  For events that are sufficiently nearby ($z\lesssim0.1$), almost all long GRBs have proven this expectation true.  There are some exceptions to this, e.g., GRBs 230307A \citep{Gillanders2023, Levan230307a, Yang2024}, 211211A \citep{ Rastinejad2022, Troja2022, Gompertz2023}, 111005A \citep{Wang2017, Michalowski2018, Tanga2018}, 060614A \citep{Gehrels2006, Fynbo2006, DellaValle2006, GalYam2006}, and 060605A \citep{Fynbo2006, Ofek2007}, which may arise from merger events.  The first long GRB to have an observed SN was GRB\,980425/SN\,1998bw \citep[$z = 0.0085$, though some now consider this GRB an X-ray flash, ][]{Galama1998, Ghisellini2006, Virgili2009}.  SN\,1998bw has extensive multi-wavelength photometry and spectroscopy to late times ($t\sim 400$ rest-frame days post-trigger) that are used to model other SNe associated with GRBs (GRB-SNe).

In the case of GRB\,221009A, there have been several searches for an associated supernova; however many studies were only able to place upper limits on its luminosity \citep{Shrestha23, Laskar2023, Levan2023} or found weak evidence for a SN detection \citep[SN\,2022xiw,][]{deUgarte2022gcn, Maiorao2022, Fulton23, Kong2024}.  The supernova was finally confidently identified with a JWST/NIRSpec spectrum at 194 days (observer-frame) post-trigger \citep{Blanchard2024}. 

To further constrain the properties of the afterglow and SN, we continued to monitor the light curve with HST and JWST.  Here, we present late-time ($t = 185 - 345$ days, observer-frame) HST/WFC3 and JWST/ NIRCam imaging of the host and afterglow of GRB 221009A.  In Section \ref{sec:observations}, we describe the observations. In Section \ref{sec:reduction}, we describe the data reduction and photometric measurement steps.  In Section \ref{sec:discussion}, we discuss the interpretation of the SED and light curve of the afterglow.  We discuss the evidence for a constant, blue component in addition to the afterglow and our detection of a break at $t \sim 50$ days, which is consistent with an interpretation as a jet break.  We conclude with comparisons to the larger GRB sample of jet-break times and optical light curves.  All times are presented in the observer-frame, unless otherwise expressly noted.  Days since trigger are measured from the Fermi trigger of 2022 Oct 09 13:17:00 UT \citep{Lesage23}.
We use a cosmological model with $H_0 = 68\ \text{km\ s}^{-1}\ \text{Mpc}^{-1}$, $\Omega_0 = 0.31$, and $\Omega_{\Lambda} = 0.69.$ 

\section{Observations}\label{sec:observations}
\subsection{Hubble Space Telescope}

We obtained observations of the afterglow with HST/WFC3 \citep[program GO 17278, PI: Chornock;][]{ChornockHST2022} on 2023 April 12 (F110W, F160W; $\Delta t = 185.19$ days), 2023 July 13 (F814W, F110W, F160W; $\Delta t_{F814W} = 277.01$ days, $\Delta t_{F110W, F160W} = 277.11$ days), 2023 September 02 (F814W; $\Delta t = 328.21$ days), 2023 September 03 (F110W, F160W; $\Delta t =328.83$ days), and 2023 October 06 (F110W, F160W; $\Delta t = 361.93$ days).  Loss-of-tracking due to loss of guide star occurred in one of the 2023 September 03 orbits, and images were retaken on 2023 October 06.  We also include HST/WFC3/F160W observations of this source from 2023 September 11 \citep[program GO/DD 17264, PI: Levan; $\Delta t = 336.79$ days;][]{LevanHST2022}.  We combine the 2023 September 03, 2023 September 11, and 2023 October 06 data into single epoch images for F110W and F160W ($\Delta t_{F160W} = 342$ days, $\Delta t_{F110W} = 345$ days).  We further refer to the imaging from 2023 April as ``V1," 2023 July as ``V2," and 2023 September (or otherwise combined) as ``V3" (Table \ref{tbl:photometry}).

We use standard STScI software \texttt{TweakReg} and \texttt{AstroDrizzle} to align and drizzle the images. We first use \texttt{TweakReg} to align the images to within one HST pixel. In \texttt{AstroDrizzle}, we use \texttt{final\_scale = 0.065} for F110W and F160W and \texttt{final\_scale = 0.02} for F814W imaging.  The \texttt{final\_pixfrac}, the fraction by which the input pixel is shrunk before being input onto the final grid, was maximized for each image following STScI recommended guidance that the standard deviation divided by the mean of the `wht' image was less than 0.20.\footnote{\url{https://hst-docs.stsci.edu/drizzpac/chapter-6-reprocessing-with-the-drizzlepac-package/6-3-running-astrodrizzle}}

\begin{deluxetable*}{lrrrrrr}
    \tablecaption{Photometry of the afterglow of GRB 221009A. From left to right: the filter and visit, with visit designations as described in Section \ref{sec:observations}; date of observations; number of days after the trigger; measured afterglow magnitude; measured uncertainty of the afterglow magnitude; systematic uncertainty of the afterglow magnitude; and, finally, the combined uncertainty which is the quadrature sum of the measured and systematic uncertainties. The \texttt{GF} and \texttt{HP} designations refer to our choice of using \texttt{GALFIT} or \texttt{HOTPANTS}, respectively, to measure the photometry (see Sec. \ref{subsec:diffim}).  The systematic uncertainty measurement procedure is detailed in Sec. \ref{subsec:sysunc}.\\
    $^*$ These data were compiled from observations on 2023 Sep 03 and 2023 Oct 06.\\
    $^{**}$ These data were compiled from observations on 2023 Sep 03, 2023 Sep 11, and 2023 Oct 06.\\ \label{tbl:photometry}}
    \tablehead{Filter \& Visit & UT Date & $\Delta t$  & Measured Mag. &  Measured Unc. & Systematic Unc. & Combined Unc. \\
     & & [days] & [mag] & [mag] & [mag] & [mag]}
    \startdata
        \textbf{HST}\\
        \hline
        F110W V1 & 2023 Apr 12 & 185.19 & 24.95 & 0.05 & $0.09$ & $0.10$\\
        F110W V2 & 2023 Jul 13 & 277.11 & 25.68 & 0.11 & $0.10$ & $0.15$\\
        F110W V3 & 2023 Sep 19$^*$ & 345 & 26.10 & 0.13 & $0.11$ & $0.17$\\
        \hline
        F160W V1 (\texttt{GF}) & 2023 Apr 12 & 185.19 & $24.67$ & $0.08$ & $0.10$ & $0.13$\\
        F160W V1 (\texttt{HP}) & 2023 Apr 12 & 185.19 &  $24.37$ & $0.07$ & $0.10$ & $0.12$\\
        F160W V2 & 2023 Jul 13 & 277.11 & $25.01$ & $0.11$ & $0.15$ & $0.19$\\
        F160W V3 & 2023 Sep 16$^{**}$ & 342 & $25.51$ & $0.12$ & $0.24$ & $0.27$\\
        \hline
        F814W V2 & 2023 Jul 13 & 277.01 & $27.00$ & $0.05$  & $0.11$ & $0.12$\\
        F814W V3 & 2023 Sep 02 & 328.21 & $27.10$ & $0.05$  & $0.12$ & $0.13$\\
        \hline
        \textbf{JWST}\\
        \hline
        F115W V1 & 2023 Apr 22 & 194.74 & 24.93 & 0.03 & 0.05 & $0.06$\\
        F200W V1 & 2023 Apr 22 & 194.74 & 24.05 & 0.02 & 0.05 & $0.06$\\
        F277W V1 & 2023 Apr 22 & 194.74 & 23.62 & 0.04 & 0.05 & $0.06$\\
        F444W V1 & 2023 Apr 22 & 194.74 & 23.11 & 0.04 & 0.05 & $0.06$\\
        \hline
        F115W V3 & 2023 Sep 04 & 330.26 & 26.29 & 0.15 & 0.10 & $0.18$\\
        F200W V3 & 2023 Sep 04 & 330.26 & 25.22 & 0.07 & 0.10 & $0.12$\\
        F277W V3 & 2023 Sep 04 & 330.26 & 24.63 & 0.15 & 0.10 & $0.18$\\
        F444W V3 & 2023 Sep 04 & 330.26 & 24.18 & 0.14 & 0.10 & $0.17$\\
    \enddata
\end{deluxetable*}

\subsection{JWST}

\cite{Blanchard2024} present JWST observations at $\Delta t = 194.74$ days post-burst, roughly contemporaneous with HST V1 ($\Delta t = 185.19$ days, a $10$ day offset).  We present additional JWST/NIRCam imaging \citep[ID: 2784, PIs: Blanchard, Chornock, Villar;][]{BlanchardJWST2022} from 2023 Sep 04 (F115W, F200W, F227W, F444W; $\Delta t = 330.26$ days).  We follow similar reduction and analysis procedures, as presented in \cite{Blanchard2024}.  The level 2 WebbPSF models were updated since the analysis presented in \cite{Blanchard2024}, so we re-reduce the $\Delta t = 194.74$ days for consistent data reduction procedure.  This results in changes in the photometry $<$ 0.17 mag.  These measurements are listed in Table \ref{tbl:photometry}.

\section{Afterglow and Host Galaxy Photometry}\label{sec:reduction}
\subsection{HST}
\subsubsection{\texttt{GALFIT} Modeling}
\begin{figure*}
    \centering
    \includegraphics[width=0.81\textwidth]{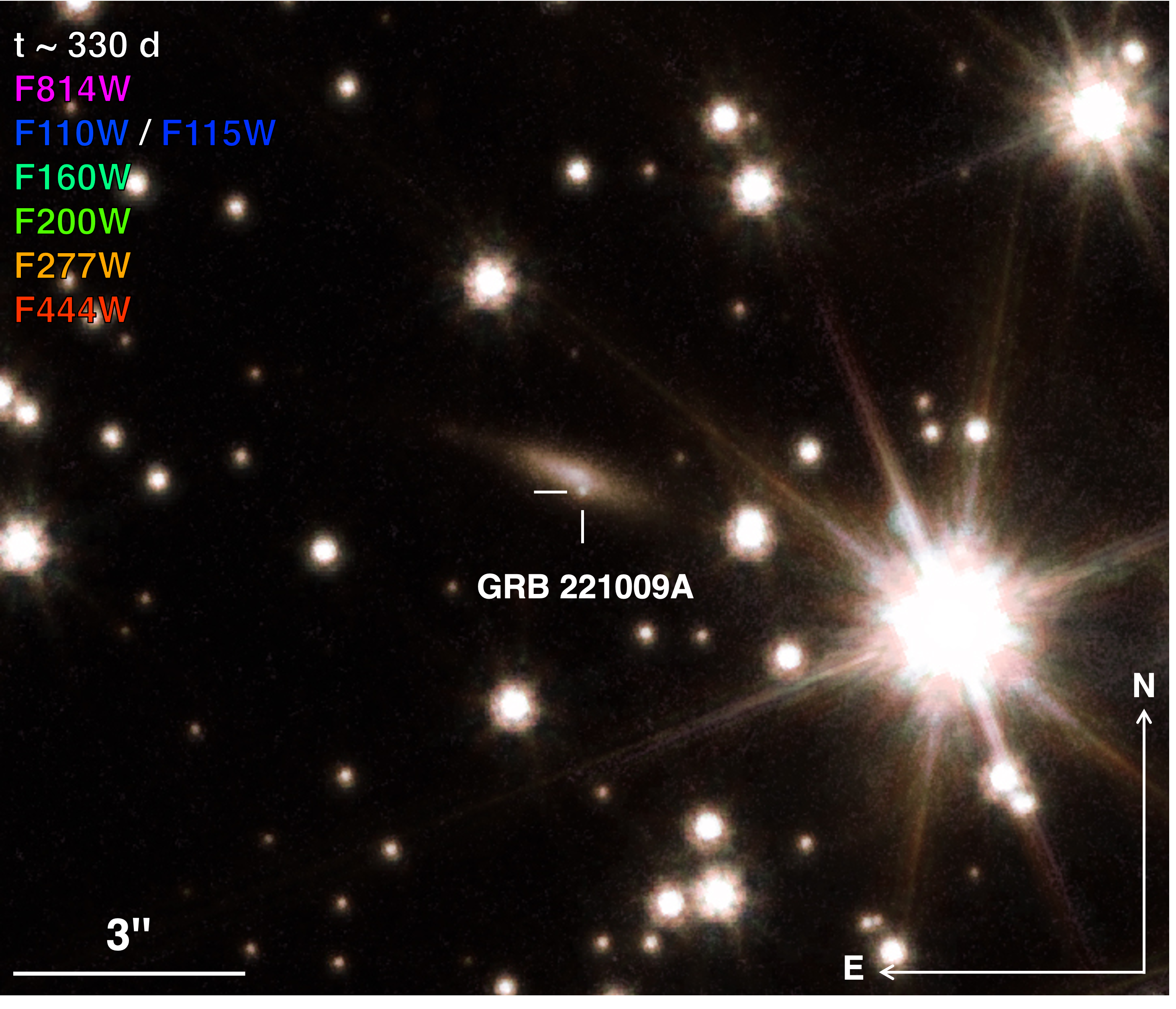}
    \includegraphics[width=0.4\textwidth]{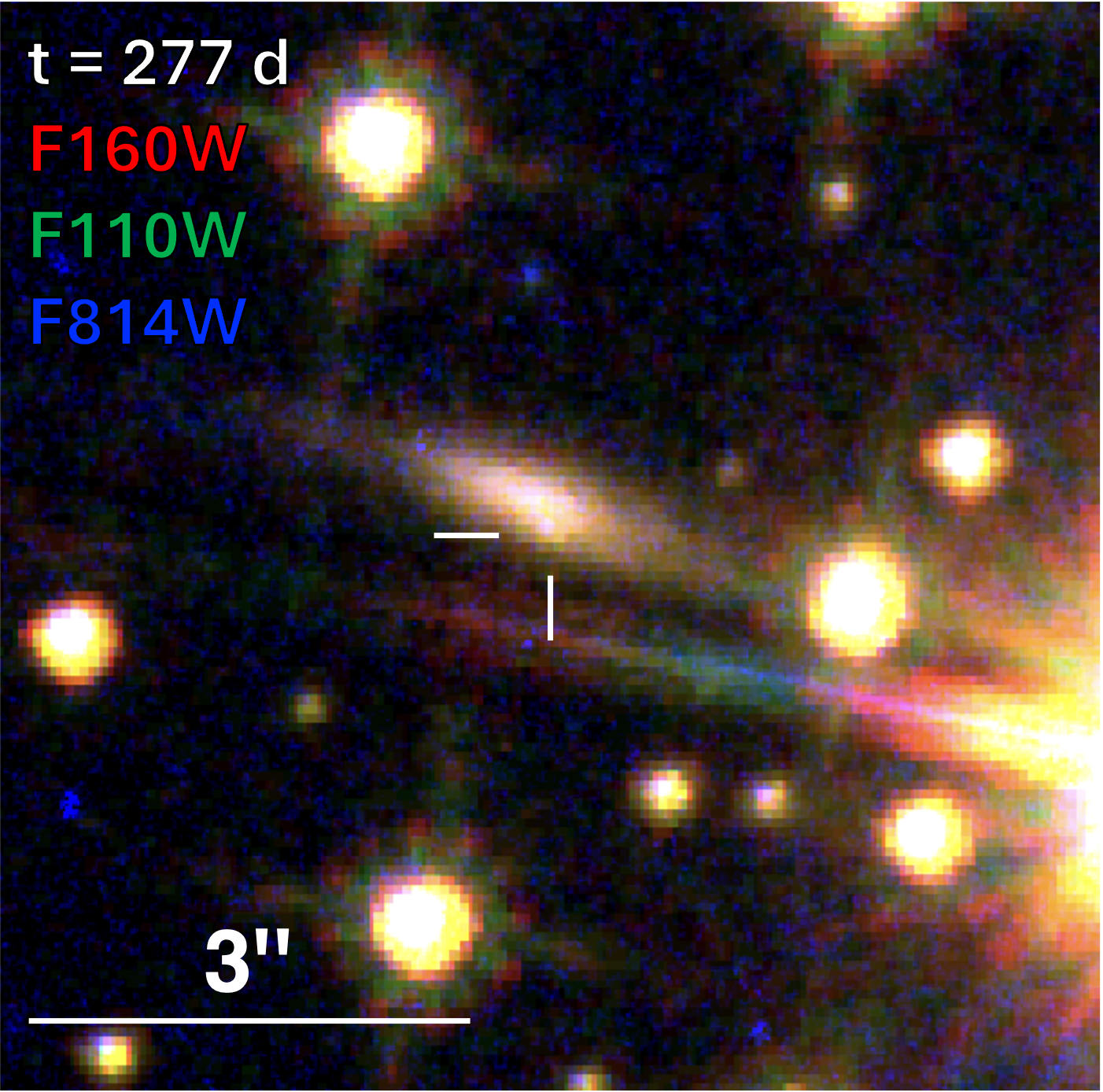}
    \includegraphics[width=0.4\textwidth]{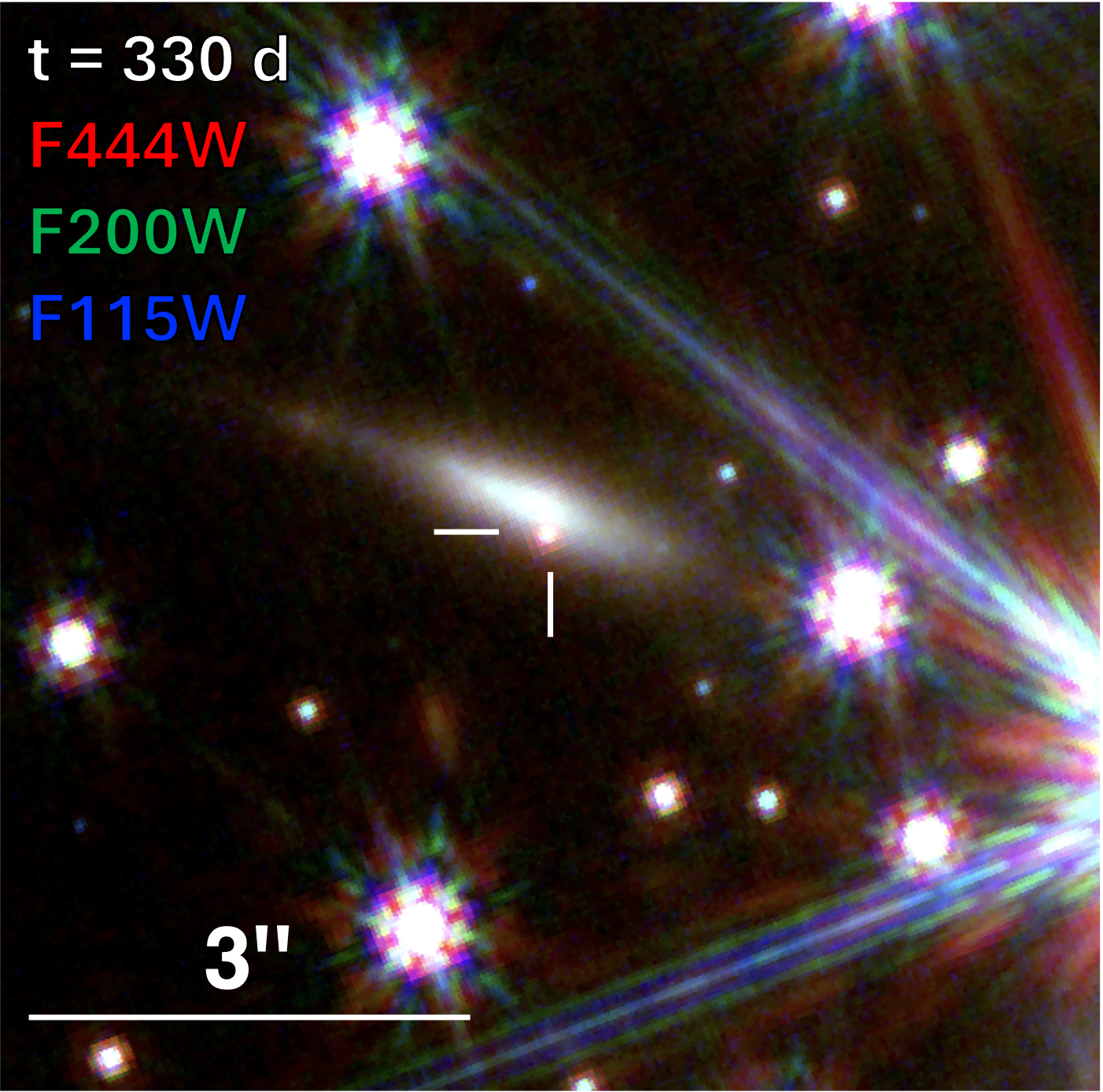}
    
    \caption{\textbf{Top:} A 15\arcsec\ HST/JWST V3 ($t \sim 330$ days) composite image centered at the afterglow. The position is marked with white cross hairs. In all panels, date of imaging, color, and filter are listed in the top left corner. North is up, and East is to the left. \textbf{Lower left:} A V2 HST composite image.  The afterglow is blended with the host galaxy and there are several diffraction spikes nearby.  \textbf{Lower right:} A V3 JWST composite image.  The afterglow is clearly detected.}
    \label{fig:composite}
\end{figure*}

We use \texttt{GALFIT} \citep{Peng2010} on each of the drizzled images to measure the magnitude of the afterglow and determine a host galaxy model.  We start by constructing a PSF for each of our images. We use a custom routine utilizing the \texttt{astropy} package \texttt{EPSFBuilder} \citep{larry_bradley_2023_7946442} and \texttt{Source Extractor} \citep{SourceExtractor} to construct a PSF model for each image. We use approximately 40 stars to construct each PSF, with each star being hand selected to avoid stars with close companions or contamination from diffraction spikes (Figure \ref{fig:composite}).  We attempted to use the same stars in each image, however, different positional angles across visits changed the orientation of diffraction spikes and total field of view in each image. We use \texttt{Source Extractor} to determine the centroid of each star to ensure the cutouts used by \texttt{EPSFBuilder} are well-centered.  In visual examinations of the output PSF, we find the best choices for the \texttt{oversampling} and \texttt{maxiters} parameters are 1 and 10, respectively, across all images.

The field around our afterglow and host galaxy is well populated with stars.  There are several stellar sources nearby that have bright diffraction spikes contaminating our source (see Figure \ref{fig:composite}).  These stars also contribute scattered light that complicates making robust measurements of the background, and hence of the  afterglow and host galaxy.  For our F110W and F160W imaging, we opt to use a fitting region for \texttt{GALFIT} that includes four additional sources to the Southeast of the host galaxy.  This allows us to maximize the number of background pixels available to \texttt{GALFIT} while excluding bright, saturated sources that \texttt{GALFIT} would be unable to model well.  Our results are not sensitive to the details of the crop region, as the final region was selected by minimizing the change in recovered afterglow magnitudes, assuming a constant galaxy model. For our F814W imaging, since the plate scale ($0.02'' = 1$ pix) is much smaller than that of our WFC3/IR imaging ($0.065'' = 1$ pix), we use a different fitting region that does not include any additional sources.  We hold each respective fitting region constant across all visits to standardize our photometry.   

We assume a S\'ersic profile as the model of the galaxy in one visit  and hold it constant across other visits.  The brightness contrast between the underlying host galaxy and afterglow is largest in the V1 imaging, so to best separate contributions from the two sources, we use V1 imaging to measure the F110W and F160W galaxy models.  To measure the F814W host galaxy model, we use V3 imaging, since we do not have V1 imaging and there is a diffraction spike going through the galaxy in V2.  The best fitting parameters for the galaxy profiles are listed in Table \ref{tbl:galmodels}.

\begin{table}
    \begin{center}
    \caption{\texttt{GALFIT} Host Galaxy Best Fitting Parameters.  $m$ is the magnitude, $R_e$ the half-light radius; $n$ the S\'ersic index, $b/a$ the axis ratio, and $\theta$ the orientation angle. \texttt{GALFIT} reports all measurements to two decimals, so the $0.00$ uncertainty on the axis ratio (for all filters) represents an uncertainty measured $<0.005$. The F110W and F160W S\'ersic models were fit to the V1 images.  The F814W model was fit to the V3 image. \label{tbl:galmodels}}
    \vskip0.1in
    \begin{tabular}{lrrr}
    \hline
    \hline
    \textbf{Parameter} & \textbf{F110W} & \textbf{F160W} & \textbf{F814W} \\
    \hline
    $m$ [mag] & 21.92 $\pm$ 0.02 & $21.29 \pm 0.02$ & 23.14 $\pm$ 0.02 \\
    $R_{e}$ [\arcsec] & $0.70 \pm 0.02$ & $0.69 \pm 0.02$ & $0.71 \pm 0.02$ \\
    $n$ & 1.21 $\pm$ 0.05 & $1.16 \pm 0.04$ & 1.04 $\pm$ 0.03 \\
    $b/a$ & 0.23 $\pm$ 0.00 & $0.23\pm 0.00$ & 0.28 $\pm$ 0.00\\
    $\theta$ [deg] & 67.39 $\pm$ 0.42 & $67.12 \pm 0.40$ & 68.66 $\pm$ 0.37 \\
    \hline
    \end{tabular}
    \end{center}
\end{table}

With these galaxy profiles, fitting regions, and PSFs, we use \texttt{GALFIT} to measure the magnitude of the afterglow.  We allow the sky to be a free ``tilted plane" model (i.e., \texttt{dy} and \texttt{dx} are free parameters within \texttt{GALFIT}).  For the F110W and F160W imaging, the afterglow and galaxy model are determined simultaneously in the V1 imaging, and for F814W, in the V3 imaging.  The residual images were visually inspected for each reported model.  The \texttt{GALFIT} derived magnitudes and their statistical uncertainties are listed in Table \ref{tbl:photometry}.  \cite{Levan2023} also fit a S\'ersic model to the host galaxy.  They find $m_{F160W} = 20.92 \pm 0.10$ mag, which is slightly brighter than our measurement of $m_{F160W} = 21.29 \pm 0.02$ mag, but within $3\sigma$ agreement.  They also find $n_{F160W} = 1.71 \pm 0.18$ and $b/a_{F160W} = 0.22 \pm 0.01$.  Our axis ratios are in agreement, however our S\'ersic index of $1.16 \pm 0.04$ is smaller, though in $3\sigma$ agreement.  The afterglow is brighter by $\sim 4$ mag at the time of the observations in \cite{Levan2023}, which could explain the differences in our galaxy models.

\subsection{Difference Imaging}\label{subsec:diffim}

\begin{figure}
    \centering
    $\vcenter{\hbox{\includegraphics[width=0.45\textwidth]{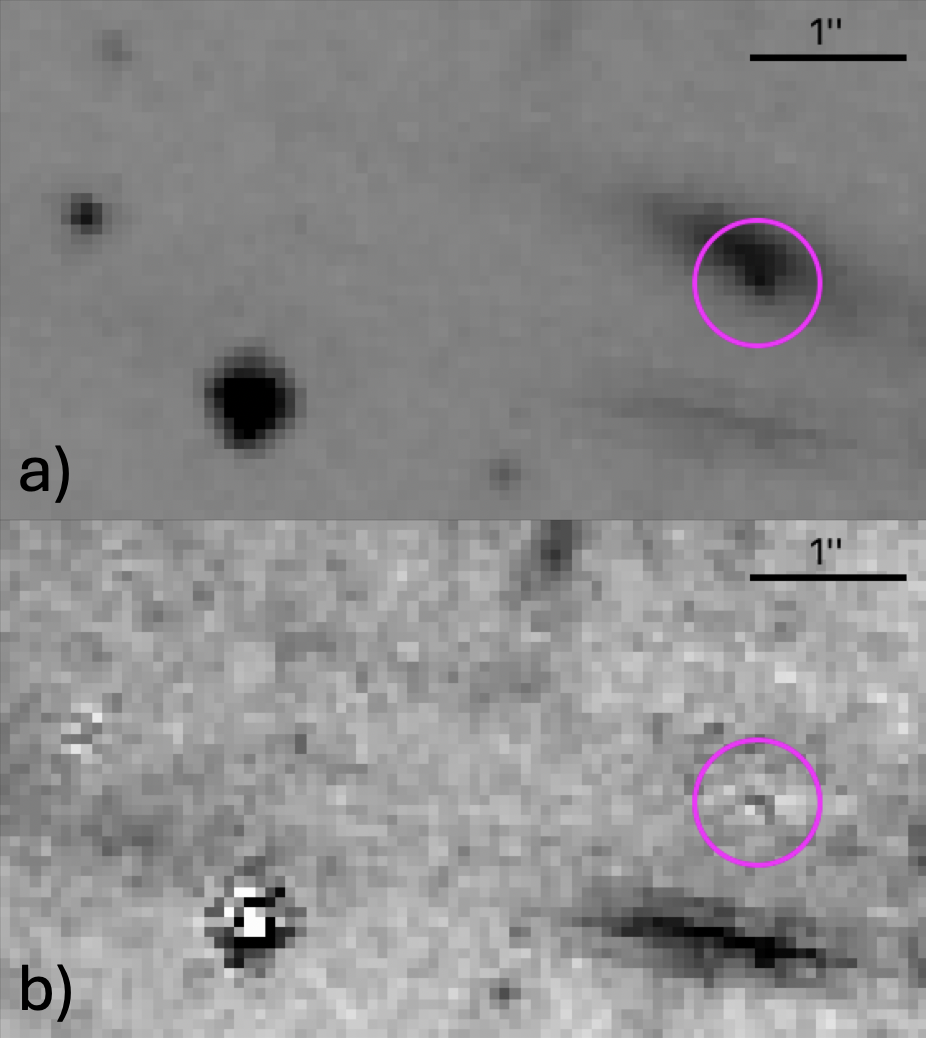}}}$ 
    
    \caption{\textbf{(a)} A crop of the V1 F160W image.  The magenta region has a radius of 0.4\arcsec\ and is centered at the location of the afterglow.  Black shows positive flux. \textbf{(b)} The \texttt{GALFIT} residual of the image above.  Scale and stretch have been modified to see fine structure. There is a slight under-subtraction of the afterglow.  North is up, and East is to the left.}
    \label{fig:v1f160wgalfit}
\end{figure}

To confirm the reliability of \texttt{GALFIT} to decompose the blended point source and host (see Figure \ref{fig:v1f160wgalfit} for an example), we perform photometry on subtracted images to measure the brightness difference of the afterglow.  We use \texttt{HOTPANTS} \citep{Becker2015} to create these subtraction images.  We have successful subtraction in the V1-V3 F160W and the V2-V3 F814W imaging which allows us to perform photometry.  In the rest, we are unable to adequately perform photometry due to scattered light.

\begin{figure}
    \centering
    $\vcenter{\hbox{\includegraphics[width=0.45\textwidth]{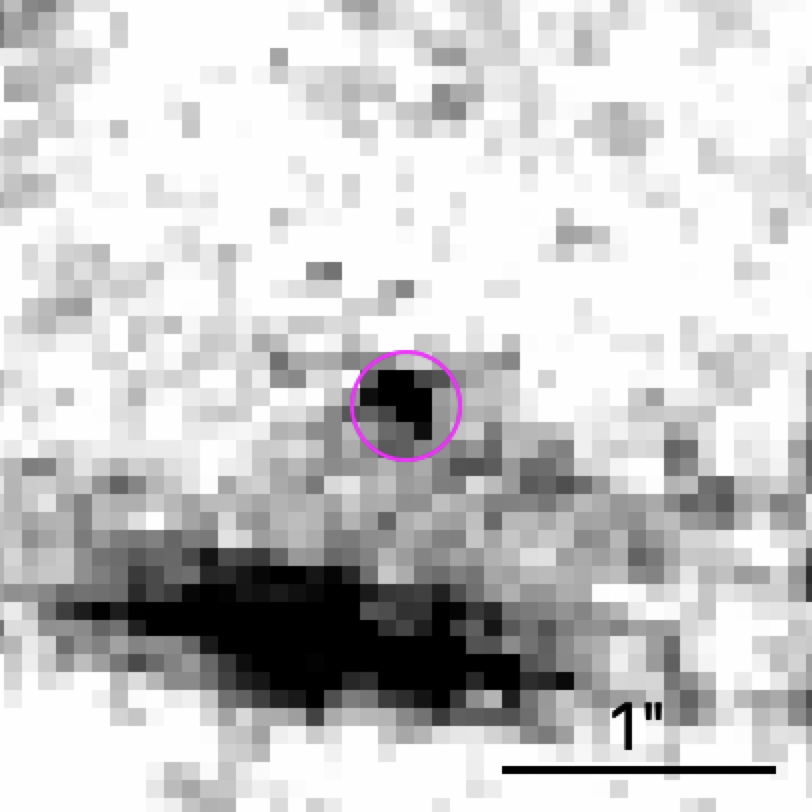}}}$ 
    $\vcenter{\hbox{\includegraphics[width=0.45\textwidth]{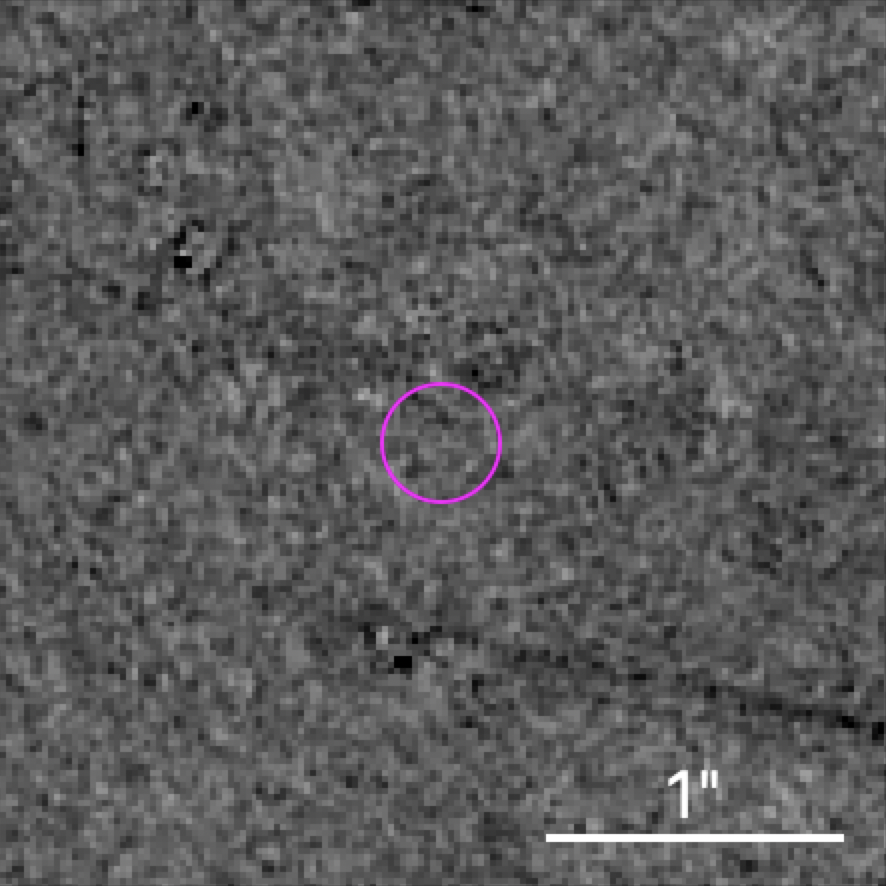}}}$
    
    \caption{\textbf{Top:} A 3\arcsec\ cutout of the subtraction image between V3 and V1 F160W. The afterglow is clearly detected in the center of the image.  There is an unsubtracted diffraction spike at the bottom center.  \textbf{Bottom:} A 3\arcsec\ cutout of the subtraction image between V3 and V2 F814W.  There is no source detected, indicating minimal fading between the two epochs.  North is up, and East is to the left.  In both panels, the magenta region has a 0.2\arcsec\ radius and is centered at the location of the afterglow.
    }
    \label{fig:HotPants}
\end{figure}

We show the usable F160W and F814W difference images in Figure \ref{fig:HotPants}.  We clearly detect a source in the F160W difference image.  We use \texttt{PhotUtils} PSF photometry \citep{larry_bradley_2023_7946442} to measure a difference flux of this source. Using the \texttt{GALFIT} V3 magnitude as an anchor, we measure a V1 F160W afterglow magnitude of $24.37 \pm 0.07$ mag.  This is consistent at the $3\sigma$ level with the \texttt{GALFIT} V1 F160W magnitude of $24.67 \pm 0.13$ mag.  As shown in Figure \ref{fig:v1f160wgalfit}, there is a slight under-subtraction of the afterglow in the \texttt{GALFIT} residual. For this reason, we defer to the \texttt{HOTPANTS} magnitude for our analysis of the V1 F160W photometry, though we find that our results are not sensitive to this choice (see Sec. \ref{subsec:mwm}).  Furthermore, we use the \texttt{HP} or \texttt{GF} shorthand to designate if we use the  \texttt{HOTPANTS} or \texttt{GALFIT} measurement, respectively.  There is no source visible at the location of the afterglow in the F814W residual image, and we confirm a non-detection to a $3\sigma$-confidence limit using aperture photometry.

\subsection{Systematic Uncertainty Estimation}\label{subsec:sysunc}
\texttt{GALFIT} is known to underestimate the uncertainty in all parameters \citep{Peng2010}.  We therefore consider the reported uncertainty as a ``partial" uncertainty of the measurement.  To estimate the systematic uncertainty, we perform a custom source injection routine using \texttt{GALFIT} to recover the magnitude. We inject $\sim 600$ point sources along an ellipse at the distance of the afterglow from the host centroid with the same $b/a$ and position angle as the S\'ersic profile (see Table \ref{tbl:galmodels}).  We do not inject sources within one PSF FWHM of the center of the afterglow. When using \texttt{GALFIT} to recover the brightness of the injected sources, we use the same image crop as was used to measure the afterglow.  We ensure that the magnitude of the injected source is the only free parameter of the injected source point source model. We apply an offset to the injected source magnitude until the median of the recovered magnitudes is within $0.01$ mag of what was measured for the afterglow. We report the one-sigma measurement of this recovered-source magnitude histogram as the systematic uncertainty.  For the JWST data, we assume 0.05 mag systematic uncertainty on the $t = 194.74$ days data and 0.10 mag systematic uncertainty on the $t = 330.26$ days data, based on experience reducing other Level 3 JWST imaging \citep{Blanchard2024}. Measurements of the systematic uncertainty are listed in Table \ref{tbl:photometry}.

\section{Discussion}\label{sec:discussion}

\subsection{HST and JWST Light Curves}
\begin{figure}
\centering
\includegraphics[width=0.45\textwidth] {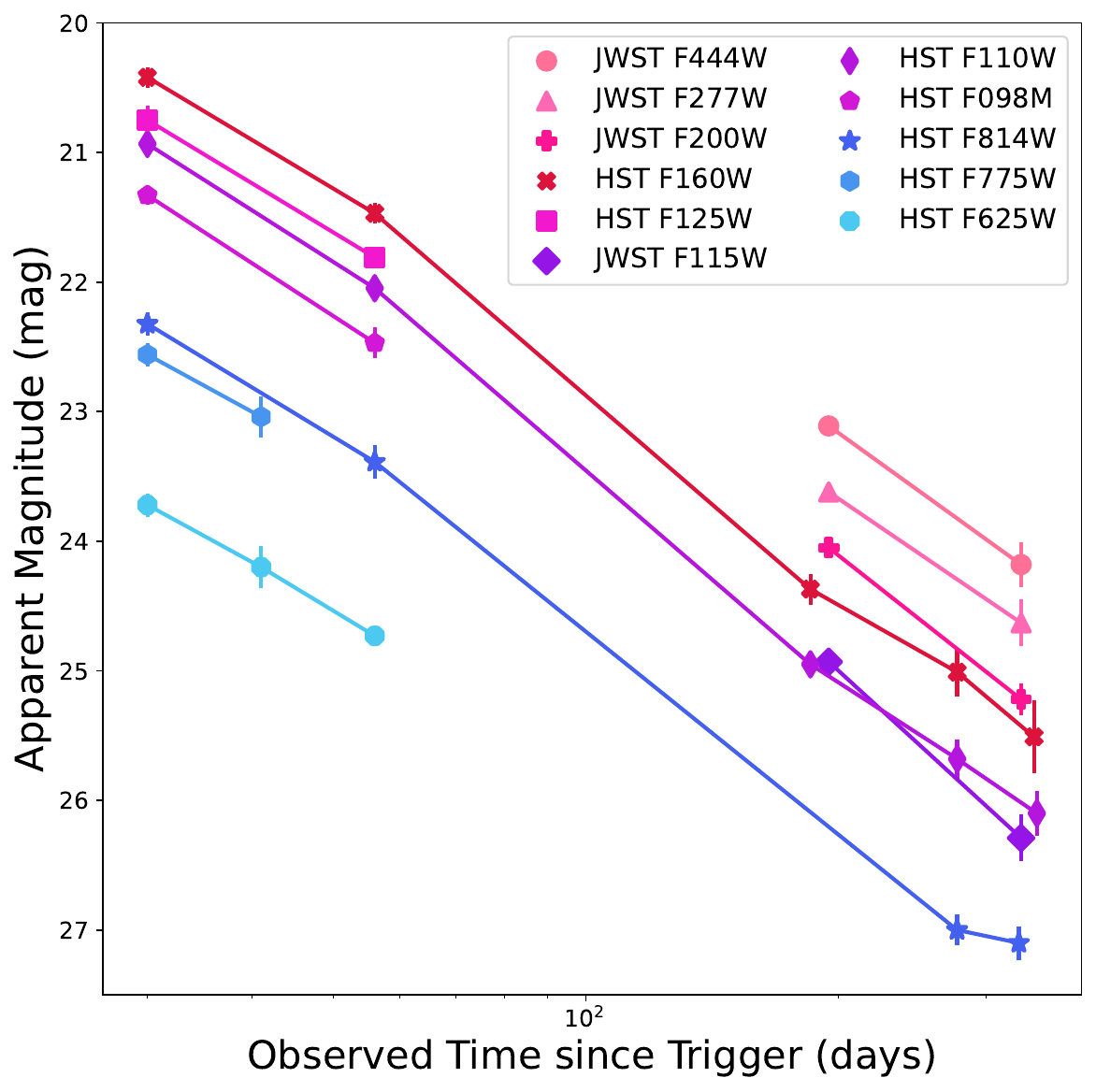}
\caption{Observed HST + JWST light curve of GRB\,221009A.  The lines connecting the data are meant to guide the eye.  The F160W measurement at $t = 185$ days (V1) is from our difference imaging.  Early HST data ($t < 60$ days) in all filters except F110W and F814W are from \cite{Levan2023}.  Early F110W and F814W data and their uncertainties were interpolated from the other filters.  Uncertainties are sometimes smaller than the marker.\label{fig:HSTLC}}
\end{figure}

We present light curves of the HST and JWST observations in Figure \ref{fig:HSTLC}.  We supplement our late-time HST observations with early HST photometry published in \cite{Levan2023}.  We cubic-spline interpolate their observations in F125W, F098M, F775W, and F625W at $t = 30\ \text{and again at}\ 56$ days post-trigger to F110W and F814W for better comparison to our observations.  One immediate observation is the flattening of the F814W light curve at $t > 277$ days.  We present detailed light curve modeling and interpretation in Section \ref{subsec:mwm}.

\subsection{SED}

We consider two dust correction models for the data.  The first law is that from \cite{Fitzpatrick99} (``F99"), while the second is from \cite{Gordon23} (``G23"), which is better calibrated for data in our observed JWST bands.  For these laws and for this source, \citet{Blanchard2024} find $A_V = 4.63\,\textrm{mag}, \ R_V = 4.24$ (F99) and $A_V = 4.37\,\textrm{mag}, \ R_V = 3.07$ (G23).  We note that these $A_V$ measurements represent the \emph{total} extinction along the line of sight.  We discuss in Section \ref{subsec:modelinterp} the effect of this choice on our results, though we find that our results are not statistically sensitive to this choice. For consistency with analysis in \cite{Blanchard2024}, we assume the F99 dust correction as the fiducial, and unless otherwise explicitly stated, it should be assumed the F99 dust correction was used.

\begin{figure}
\centering
\includegraphics[width=0.47\textwidth]{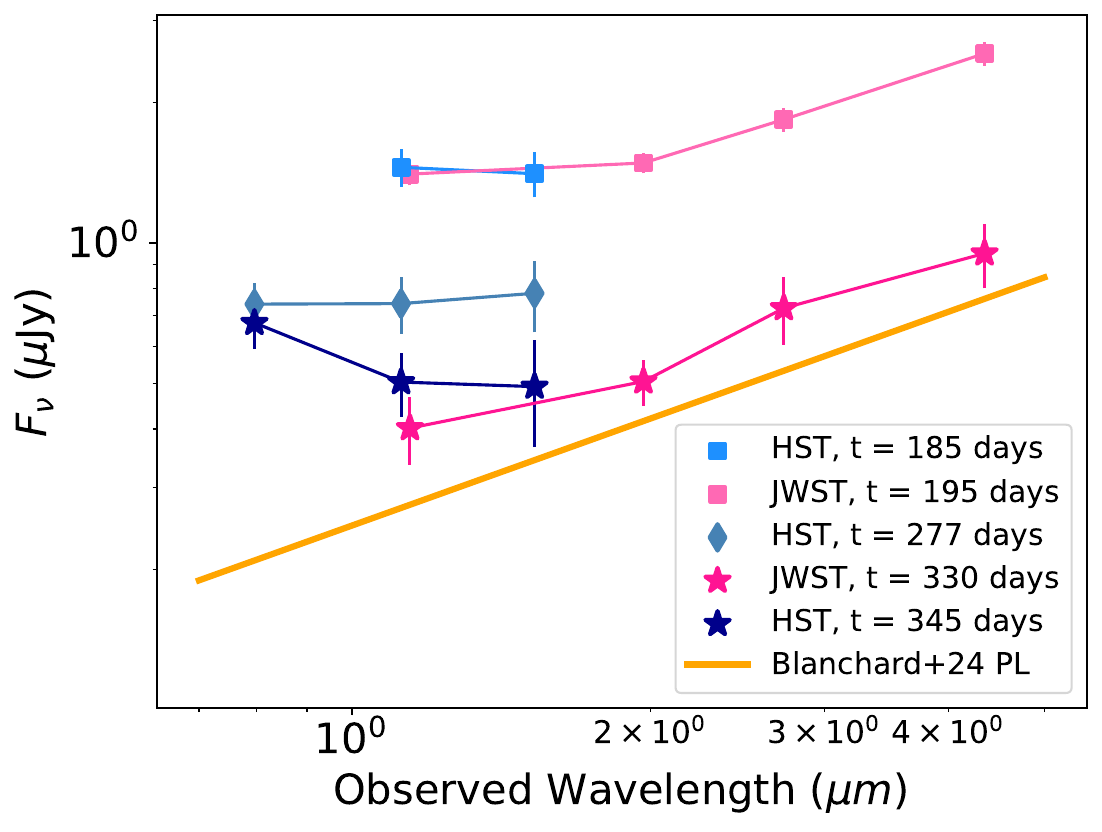}
\caption{Late-time dust-corrected HST and JWST SED of GRB\,221009A.  As in Figure \ref{fig:HSTLC}, the lines connecting the data are meant to guide the eye.  We use the \texttt{HOTPANTS} measurement for V1 F160W.  The data have been extinction corrected assuming the F99 dust model ($A_V = 4.63$ mag, $R_V = 4.24$).  The JWST photometry at $t = 195\,\rm{days}$ and the best-fit power-law model for the afterglow spectrum (orange) are from \cite{Blanchard2024} with $F_{\nu} \propto \nu^{-0.76}$.  V1 is shown with squares, V2 with diamonds, and V3 with stars. The power-law spectrum has been arbitrarily scaled. There is a clear disagreement between the power-law model and the afterglow photometry at wavelengths blueward of 2 $\mu$m at $t\sim330$ days.\label{fig:SED}}
\end{figure}

We present the SED (HST/F814W - JWST/F444W) at each epoch in Figure \ref{fig:SED}.  We have near contemporaneous HST and JWST observations at $t\sim190$ days and at $t\sim330$ days, and we find good agreement between the two data sets based on the smooth SED shape across HST/F110W - JWST/F200W.  We include in Figure \ref{fig:SED} the best-fit spectral power law ($F_{\nu} \propto \nu^{-0.76}$) for the afterglow from \cite{Blanchard2024}.  This power law shows agreement with both JWST data sets redward of $2 \mu$m, but we see evidence for a strong blue component.  The disagreement from the afterglow power-law is strongest in F814W, but is evident in all HST filters at $t\sim330$ days.  At $t\sim 330$ days, we do not expect this source to be exclusively explained by a SN similar to other GRB-associated SNe \citep{Clocchiatti2011}.  To test this expectation, we attach a $F_{\lambda} \propto \lambda^{-4}$ blackbody tail to the BVRI measurements of SN\,1998bw from \cite{Clocchiatti2011}.  We then K-correct the SN\,1998bw $f_{\lambda}$ to the distance of GRB\,221009A and calculate $f_{\nu}$ at $t = 330$ days (observer-frame) in the HST and JWST filters.  The $t=330$ day SN spectrum peaks in F814W at $0.2\,\mu$Jy, decreasing as $\lambda^{-4}$ with increasing wavelength.  To explain the blue component, the SN associated with GRB\,221009A would need to be a factor of $\sim 3$ times brighter than SN\,1998bw--strongly inconsistent with early results finding an upper limit of $\sim0.7 \times$ SN\,1998bw \citep{Gokul2023}.  A natural explanation for the blue excess is that in addition to the fading afterglow and supernova, there is some additional source--perhaps a star cluster or a light echo--underlying the GRB that is not accounted for in our host subtraction.  We explore these scenarios in Section \ref{subsec:modelinterp}.

\subsection{Light-curve  Modeling}\label{subsec:mwm}
We fit three afterglow decay models to our data: (1) a single power law plus supernova,

\begin{equation}
\begin{split}
f_{\nu}(t) =  t^{-\alpha_1} + S_{SN} \times F_{SN\,1998bw},
\end{split}
\end{equation}

\noindent
(2) a broken power-law plus supernova, 

\begin{equation}
\begin{split}
f_{\nu}(t) & =  2^{1/s}\times F_{t_{brk}}\times \Bigg(\Bigg(\frac{t}{t_{brk}}\Bigg)^{\alpha_1 s} +\Bigg (\frac{t}{t_{brk}}\Bigg)^{\alpha_2 s}\Bigg)^{-1/s} \\
& +\ S_{SN} \times F_{SN\,1998bw},
\end{split}
\end{equation}

\noindent
and (3) a broken power-law plus supernova plus constant source 

\begin{equation}
\begin{split}
f_{\nu}(t) & =  2^{1/s}\times F_{t_{brk}}\times \Bigg(\Bigg(\frac{t}{t_{brk}}\Bigg)^{\alpha_1 s} +\ \Bigg (\frac{t}{t_{brk}}\Bigg)^{\alpha_2 s}\Bigg)^{-1/s} \\
& +\ S_{SN} \times F_{SN\,1998bw} +\ F_{c},
\end{split}
\end{equation}

\noindent
where $t$ is the time in days, $S_{SN}$ is the flux scaling of SN\,1998bw, $F_{SN\,1998bw}$ is the flux of SN\,1998bw, $t_{brk}$ is the break time in days, $F_{t_{brk}}$ is the flux at time $t = t_{brk}$, $\alpha_1$ is the pre-break (or only) slope, $\alpha_2$ is the post-break slope, $s$ is the smoothing parameter, and $F_{c}$ is the flux of the constant source.  While we are unable to fit for the smoothing parameter with our data set, because the physical processes governing the two slopes is the same, we expect a relatively smooth transition.  We elect to use $s=3$, however our results are not significantly sensitive to this choice.

For the SN component, we assume a SN 1998bw light curve with allowance for time stretching. We start with the host-extinction corrected BVRI (rest-frame) data from \cite{Clocchiatti2011} and attach a $f_{\lambda}\propto \lambda^{-4}$ blackbody tail redward to estimate $f_{\lambda}$.  We then K-correct the SN\,1998bw $f_{\lambda}$ to the distance of GRB\,221009A and calculate light curves in each of the filters of interest.

\begin{figure*}
\centering
\includegraphics[width = 0.49 \textwidth]{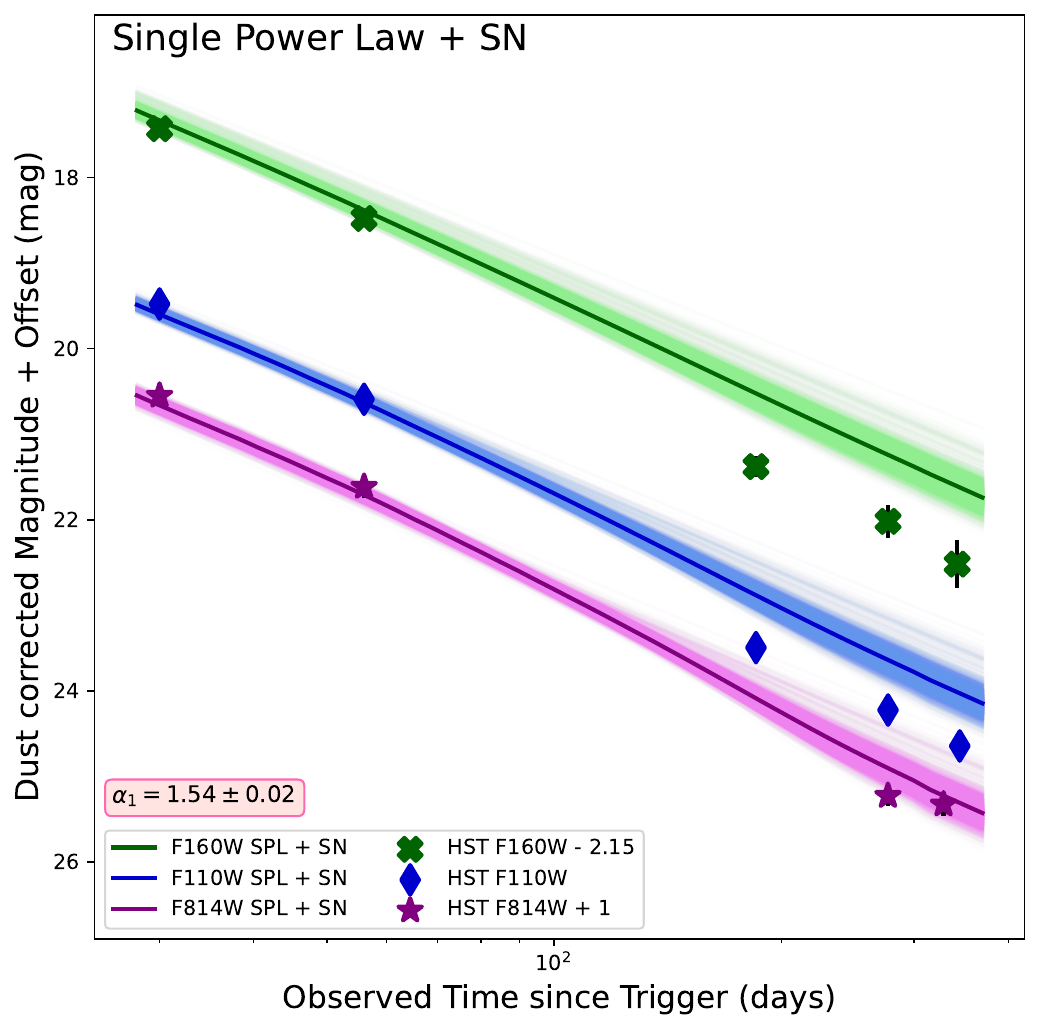}
\includegraphics[width = 0.49 \textwidth]{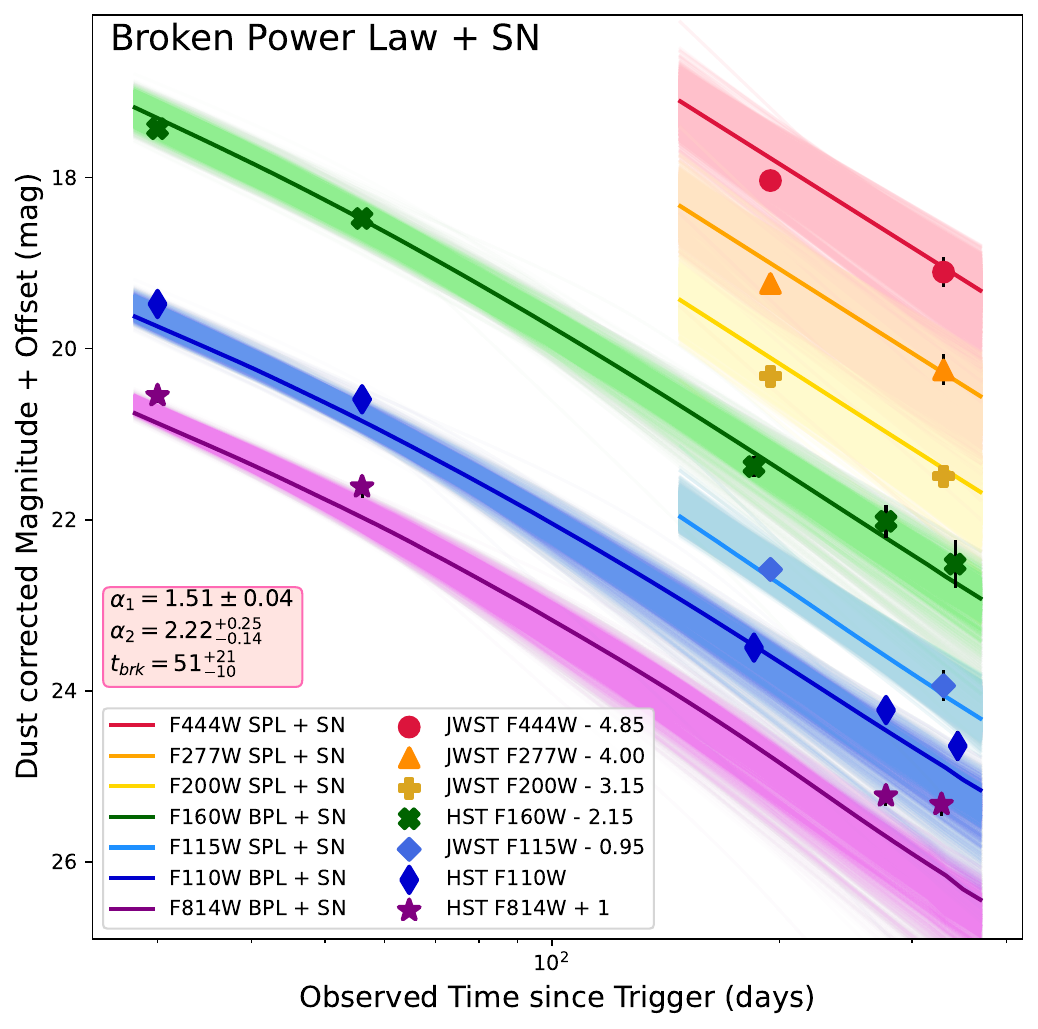}
\includegraphics[width = 0.49 \textwidth]{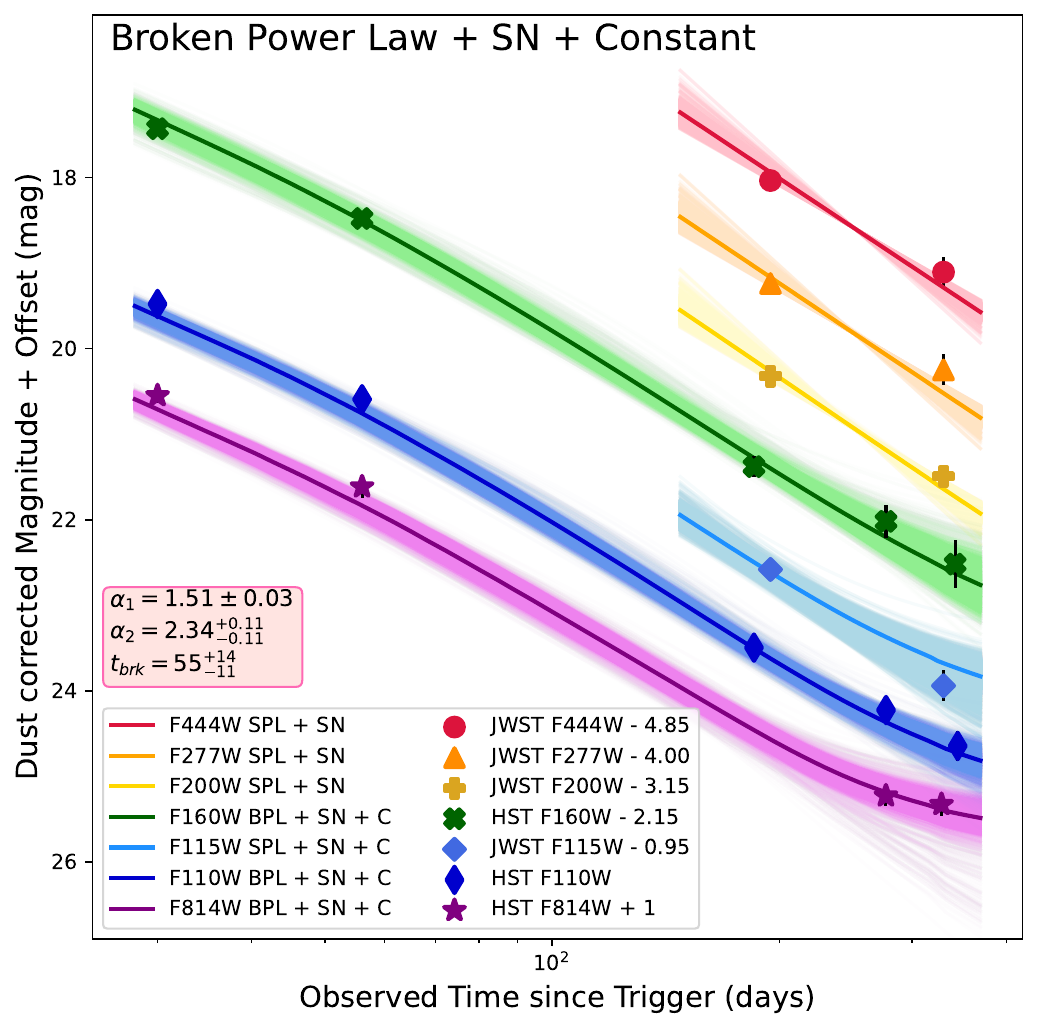}
\caption{\textbf{Top Left:} Single power-law fits to early-time data from \cite{Shrestha23}, \cite{Fulton23}, and \cite{Levan2023} extended to late times.  3060 ($68\% \text{ of } 4500$) random pulls from each posterior are shown as the lighter shaded lines to show the $1\sigma$ uncertainty of the fit.  HST observations from \cite{Levan2023} and those published here are shown as the connected error-bars.  These observations have been dust-corrected assuming the F99 dust law ($A_V = 4.63$ mag, $R_V = 4.23$) and have been artificially offset. We use the \texttt{HOTPANTS} measurement for V1 F160W.   These observations and choices are consistent across all panels. There is clear disagreement between the single power-law plus a SN model extrapolations and the late-time observations in all filters, F160W especially. \textbf{Top Right:}  The models fit to the data are now a broken power law + SN, and the JWST data have been added.  \textbf{Bottom:} The models fit to the data are now a broken power law + SN + constant.  The fitting procedures for all panels are described in Section \ref{subsec:mwm}, and all fitting parameters are listed in Table \ref{tbl:emceeresults}.  Uncertainty bars are sometimes smaller than the symbols.
\label{fig:3panel1}}
\end{figure*}

We fit the single power-law plus scaled SN (``SPL + SN'') to the early data \citep[t = 1--56 days, $riz$, $rizy_{PS}$, and HST; ][respectively]{Shrestha23, Fulton23, Levan2023} using a Markov Chain Monte Carlo (MCMC) sampler with \texttt{emcee} \citep{emcee2013} and extrapolate to later times.  We fit for the magnitude at $\Delta t = 10$ days in F110W and apply $f_{\nu} \propto \nu^{-0.76}$ \citep{Blanchard2024} to calculate the magnitude in other filters. We assume the same $\alpha_1$ for all filters. We run 32 walkers for 5000 steps and discard the first 500 steps as burn-in.  We use uniform priors on $\alpha$ from 1.0 to 3.5, $m_{fit, F110W}$ from 10 to 40 mag, t\_stretch of 0.5 to 3.0, and $S_{SN}$ of 0 to 10.  Shown in the top left panel of Figure \ref{fig:3panel1}, the extrapolation of the PL model is a poor fit to the later ($t > 185$ days) data.  We report the results of our fitting in Table \ref{tbl:emceeresults}.

We repeat our MCMC \texttt{emcee} fitting now assuming single power-laws for the $riz$, $rizy_{PS}$, and JWST data and a broken power-law for the full HST data with the addition of a SN in all filters.  We refer to this model as ``BPL + SN.''  The choice to use single power-laws for the ground and JWST data is because they do not cover the break window and therefore are not able to probe $t_{brk}$ or the opposing slope.  We again fit for the magnitude at $t_{brk}$ in F110W and apply $f_{\nu} \propto \nu^{-0.76}$ to calculate the magnitude in other filters.  We assume the same $\alpha_1$ and $\alpha_2$ for all filters.  We assume uniform priors on $\alpha_1$ of 1.0 to 2.0, $\alpha_2$ of 2.0 to 5.0, $\log(t_{brk})$ of 1.00 to 2.27 log(days), $m_{t_{brk}, F110W}$ of 20 to 30 mag, and again t\_stretch of 0.5 to 3.0 and $S_{SN}$ of 0 to 10.  We assume $\alpha_1$ for the slopes of the $riz$ and $rizy_{PS}$ power-laws and $\alpha_2$ for the slopes of the JWST power-laws.  We show our fits to the HST and JWST data in the top right panel of Figure \ref{fig:3panel1}. This model is noticeably better than the extrapolation of the single power-law, however there is consistent under-prediction of the entire F814W data set and most of the late-time F110W data.  We report the results of our fitting in Table \ref{tbl:emceeresults}.

The next fit to the data is the addition of a constant source to the broken power-law plus SN model described above (``BPL + SN + C'').  This choice is motivated by the under-predictions from the broken power-law plus SN modeling and the deviation in the SED from the assumed spectral power-law ($f_{\nu} \propto \nu^{-0.76}$) from \cite{Blanchard2024}, as shown in Figure \ref{fig:SED}.  We again calibrate to the F110W data set assuming $f_{\nu} \propto \nu^{-0.76}$ and assume the same $\alpha_1$ and $\alpha_2$ for all filters.  We assume the same uniform priors as before. We fit for the constant-source magnitude in each HST filter and JWST F115W, with a prior on each $m_{c}$ of 24 to 29 mag.  We assume the same constant for F110W and F115W due to their near equivalent effective wavelengths and to reduce the total number of parameters.  We are not able to probe the contribution of a late-emerging constant with the early ground data, and the V3 SED shows no evidence for an additional component in F200W, F277W, and F444W, so we set the constant in these filters to have 0 flux.  We show our fits in Figure \ref{fig:3panel1}, and  we report the results of our fitting in Table \ref{tbl:emceeresults}. 

The final fit to the data is a broken power-law plus constant with no consideration of a SN contribution (``BPL + C").  While this is an unphysical model, given the spectroscopic detection in \cite{Blanchard2024}, this can be used to test the sensitivity of our model to the inclusion of a SN component.  We repeat the same priors as before, and we assume the \texttt{HOTPANTS} measurement of V1 F160W.  We also repeat the same assumptions about which filters have a constant with 0 flux.  We report the results of our fitting in Table \ref{tbl:emceeresults}.

\begin{deluxetable*}{lrrrrrrrr}
    \tablecaption{\texttt{emcee} results following the procedure detailed in Sec. \ref{subsec:mwm}.  The normalization magnitude for the single power-law (SPL) + SN is $m_{F110W}$ at time t = 1 day. The normalization magnitude for the broken power-law (BPL) + SN, and BPL + SN + Constant (C), and BPL + C is $m_{F110W}$ at time t = $t_{brk}$ log(days).  The choice of V1 F160W for the BPL models is indicated in the `SN Scale' column.  F99 refers to the choice of dust correction where we use the \cite{Fitzpatrick99} law with $A_V = 4.63$ mag, $R_V = 4.23$, and G23 refers to the choice of the \cite{Gordon23} law with $A_V = 4.37$ mag, $R_V = 3.07$. `HP' (`GF') is meant to refer to the \texttt{HOTPANTS} (\texttt{GALFIT}) choice of V1 F160W.  \label{tbl:emceeresults}}
    \tablehead{SN Scale ($S_{SN}$) & $t_{\text{stretch}}$ & $t_{brk}$ & $\alpha_1$ & $\alpha_2$ & $m_{\text{norm}}$ & $C_{F160W}$ & $C_{F110W}$& $C_{F814W}$\\ 
    & & [log days] &  &  & [mag] & [mag] & [mag] & [mag]} 
    \startdata
        \hline
        \multicolumn{3}{l}{\textbf{SPL + SN}}\\
        \hline
        \texttt{F99 HP}: $1.96_{-0.38}^{+0.37}$ & $0.52_{-0.02}^{+0.04}$ & - & $1.57_{-0.02}^{+0.02}$ & - & $17.98_{-0.04}^{+0.04}$ & - & - & -\\
        \texttt{G23 HP}: $0.71_{-0.34}^{+0.31}$ & $0.57_{-0.06}^{+0.28}$ & - & $1.49_{-0.02}^{+0.02}$ & - & $18.04_{-0.04}^{+0.04}$ & - & - & -\\
        \hline
        \multicolumn{3}{l}{\textbf{BPL + SN}}\\
        \hline
        \texttt{F99 HP}: $1.13_{-0.83}^{+0.54}$ & $0.56_{-0.06}^{+0.99}$ & $1.71_{-0.09}^{+0.15}$ & $1.51_{-0.04}^{+0.04}$ & $2.22_{-0.13}^{+0.25}$ & $20.81_{-0.42}^{+0.73}$ & - & - & -\\
        \texttt{F99 GF}: $1.18_{-0.80}^{+0.60}$ & $0.56_{-0.05}^{+0.80}$ & $1.70_{-0.08}^{+0.15}$ & $1.52_{-0.04}^{+0.05}$ & $2.21_{-0.12}^{+0.31}$ & $20.80_{-0.40}^{+0.70}$ & - & - & -\\
        \texttt{G23 HP}: $0.16_{-0.06}^{+0.11}$ & $1.99_{-0.53}^{+0.60}$ & $1.70_{-0.06}^{+0.07}$ & $1.44_{-0.02}^{+0.02}$ & $2.41_{-0.12}^{+0.14}$ & $20.72_{-0.26}^{+0.30}$ & - & - & -\\
        \texttt{G23 GF}: $0.15_{-0.07}^{+0.15}$ & $1.98_{-1.07}^{+0.67}$ & $1.68_{-0.07}^{+0.08}$ & $1.44_{-0.02}^{+0.02}$ & $2.43_{-0.19}^{+0.20}$ & $20.65_{-0.26}^{+0.34}$ & - & - & -\\
        \hline
        \multicolumn{3}{l}{\textbf{BPL + SN + C}}\\
        \hline
        \texttt{F99 HP}: $1.47_{-0.36}^{+0.39}$ & $0.55_{-0.04}^{+0.06}$ & $1.74_{-0.11}^{+0.11}$ & $1.51_{-0.03}^{+0.03}$ & $2.34_{-0.11}^{+0.11}$ & $21.00_{-0.49}^{+0.52}$ & $25.54_{-0.24}^{+0.35}$ & $24.84_{-0.18}^{+0.24}$ & $26.13_{-0.65}^{+1.22}$\\
        \texttt{F99 GF}: $1.30_{-0.70}^{+0.44}$ & $0.57_{-0.06}^{+0.21}$ & $1.64_{-0.16}^{+0.11}$ & $1.50_{-0.04}^{+0.04}$ & $2.28_{-0.15}^{+0.15}$ & $20.56_{-0.70}^{+0.54}$ & $25.60_{-0.27}^{+0.55}$ & $24.87_{-0.27}^{+0.41}$ & $26.43_{-0.85}^{+1.40}$\\
        \texttt{G23 HP}: $0.36_{-0.24}^{+0.33}$ & $0.71_{-0.17}^{+0.54}$ & $1.75_{-0.08}^{+0.07}$ & $1.45_{-0.02}^{+0.02}$ & $2.45_{-0.10}^{+0.10}$ & $20.95_{-0.38}^{+0.33}$ & $25.59_{-0.24}^{+0.44}$ & $25.22_{-0.23}^{+0.32}$ & $25.97_{-0.60}^{+1.27}$\\
        \texttt{G23 GF}: $0.29_{-0.19}^{+0.34}$ & $0.74_{-0.23}^{+0.77}$ & $1.72_{-0.12}^{+0.10}$ & $1.45_{-0.02}^{+0.03}$ & $2.41_{-0.15}^{+0.17}$ & $20.82_{-0.50}^{+0.46}$ & $25.72_{-0.34}^{+0.61}$ & $25.34_{-0.32}^{+0.45}$ & $26.40_{-0.90}^{+1.44}$\\
        \hline
        \multicolumn{3}{l}{\textbf{BPL + C}}\\
        \hline
        \texttt{F99 HP}: - & - & $1.61_{-0.09}^{+0.09}$ & $1.43_{-0.02}^{+0.02}$ & $2.35_{-0.09}^{+0.10}$ & $20.12_{-0.38}^{+0.36}$ & $25.48_{-0.31}^{+0.55}$ & $24.84_{-0.26}^{+0.31}$ & $26.63_{-1.00}^{+1.43}$\\
        \texttt{G23 HP}: - & - & $1.73_{-0.07}^{+0.07}$ & $1.42_{-0.02}^{+0.02}$ & $2.43_{-0.09}^{+0.09}$ & $20.73_{-0.27}^{+0.27}$ & $25.56_{-0.27}^{+0.43}$ & $25.19_{-0.23}^{+0.27}$ & $26.32_{-0.74}^{+1.29}$\\
    \enddata
\end{deluxetable*}

We perform $\chi^2$ tests for these data and models.  We set an uncertainty floor of $0.10$ mag.  For the best-fit SPL + SN model, we measure a $\chi^2$ of 267 with 124 degrees of freedom (d.o.f.). For the best-fit BPL + SN model, we measure a statistic of 249 with 129 d.o.f.  For the best-fit BPL + SN + C model, we measure a statistic of 211 with 126 d.o.f. For the best-fit BPL + C model, we measure a statistic of 217 with 128 d.o.f.  The SPL + SN model is only compared to the ground and HST data, whereas the other models are also compared to the JWST data, so has a slightly differently defined number of degrees of freedom.  From the BPL + SN to BPL + SN + C, we see a decrease in the $\chi^2$ of 38 with a decrease of only 3 d.o.f.  Conversely, the $\chi^2$ change from the BPL + SN + C to the BPL + C is an increase of 7 with a decrease of 2 d.o.f.  This implies that the constant is strongly preferred while the supernova component is weakly preferred.

There is some disagreement on the use of reduced $\chi^2$ tests when assuming non-linear models \citep[e.g.,][]{Andrae2010}, and so we additionally calculate an Akaike Information Criterion \citep[AIC,][]{AkaikeAIC} and Bayesian Information Criterion \citep[BIC,][]{SchwarzBIC} for each afterglow model.  As shown in Table \ref{tbl:stats}, we are able to statistically rule out both the SPL + SN extrapolations and BPL + SN models with $\Delta AIC$, $\Delta BIC > 10$ \citep{Raftery1995, Burnham2004}.  We also find that our results are not sensitive to our choice of the V1 F160W measurement.  We proceed in the next section with interpretation of the BPL + SN + C model.

\begin{deluxetable}{lrrrr}
    \tablecaption{Statistical results for the considered transient models. $\Delta \text{AIC} = \text{AIC}_{model} - \text{AIC}_{min}$. $\Delta$BIC is similarly defined. `F99' refers to the choice of dust correction where we use the \cite{Fitzpatrick99} law with $A_V = 4.63$ mag, $R_V = 4.23$, and `G23' refers to the choice of the \cite{Gordon23} law with $A_V = 4.37$ mag, $R_V = 3.07$.\label{tbl:stats}}
    \tablehead{\colhead{Model} & \colhead{AIC} & \colhead{$\Delta$AIC} & \colhead{BIC} & \colhead{$\Delta$BIC}} 
    \startdata
        \texttt{HOTPANTS} F160W - F99\\
        \hline
        SPL + SN& $-66$ & $167$ & $-49$ & $149$\\
        BPL + SN & $-202$ & $31$ & $-176$ & $22$\\
        BPL + SN + C & $-233$ & $0$ & $-198$ & $0$\\
        BPL + C & $-219$ & $14$ & $-189$ & $8$\\
        \hline
        \texttt{GALFIT} F160W - F99\\
        \hline
        SPL + SN& $-38$ & $189$ & $-20$ & $171$\\
        BPL + SN& $-201$ & $25$ & $-174$ & $17$\\
        BPL + SN + C & $-226$ & $0$ & $-191$ & $0$\\
        \hline
        \texttt{HOTPANTS} F160W - G23\\
        \hline
        SPL + SN& $66$ & $321$ & $83$ & $309$\\
        BPL + SN & $-241$ & $14$ & $-215$ & $11$\\
        BPL + SN + C & $-255$ & $0$ & $-220$ & $6$\\
        BPL + C & $-255$ & $0$ & $-226$ & $0$\\
        \hline
        \texttt{GALFIT} F160W - G23\\
        \hline
        SPL + SN& $97$ & $343$ & $114$ & $325$\\
        BPL + SN& $-235$ & $11$ & $-209$ & $2$\\
        BPL + SN + C & $-246$ & $0$ & $-211$ & $0$\\
    \enddata
\end{deluxetable}

\subsection{Model Interpretation}\label{subsec:modelinterp}
In the broken power-law plus supernova plus constant model, using the image subtraction measurement for V1 F160W, we find $\alpha_1 = 1.51 \pm 0.03$; $\alpha_2 = 2.34 \pm 0.11$; $t_{brk} = 10^{1.74 \pm 0.11}$ days; a SN\,1998bw flux scaling, $S_{SN} = 1.47^{+0.39}_{-0.36}$; a SN\,1998bw time stretch factor, t\_stretch $= 0.57^{+0.06}_{-0.04}$; and dust-corrected constant source magnitudes of $m_{F814W} = 26.13^{+1.22}_{-0.65}$ mag, $m_{F110W/F115W} = 24.84^{+0.24}_{-0.18}$ mag, $m_{160W} = 25.54^{+0.35}_{-0.24}$ mag. All parameter measurements for all models and choices of V1 F160W are in Table \ref{tbl:emceeresults}.

\subsubsection{Supernova Component}
In the preferred model, we find a SN component with a SN\,1998bw flux scaling, $S_{SN}$, in 3$\sigma$ consistency with the early results \citep[e.g., ][]{Levan2023, Laskar2023, Shrestha23, Fulton23, Gokul2023} and in 2$\sigma$ consistency with the results from the JWST-focused study \citep{Blanchard2024}. This is only true, however, due to the large uncertainty on our measurement.  We ran a model assuming a BPL + C with no SN, and we find all parameters (except $\alpha_1$) in $\sim 1 \sigma$ agreement with our results from the broken power-law plus constant plus supernova fit. The ``new" $\alpha_1 = 1.43 \pm 0.02$, is, instead, only in $3\sigma$ agreement with the BPL + SN  +  C measured quantity. In a statistical AIC test, we are able to rule out this model \citep{Burnham2004}, however since our $\Delta$ BIC $<$ 10, this model is less preferred \citep{Raftery1995} rather than ruled out. In the repeated analysis assuming the G23 dust correction, for the BPL+SN+C model, we find a flux scaling factor of $\sim 0.3 \pm 0.3$, which is in $\sim 3\sigma$ consistency with the early results and the F99 results for the same model, however, this is again only true due to the large uncertainty.  Lastly, as mentioned in the previous section, the $\chi^2$ results show a weak preference for the supernova component.  We therefore consider our result of the SN\,1998bw flux scaling of $S_{SN} = 1.47^{+0.39}_{-0.36}$ as a broad upper-limit less so than a strong detection of SN\,2022xiw.

\subsubsection{Jet Break}
We find support for a break at $t_{brk} = 10^{1.7 \pm 0.1}$ ($\sim 50$) days.  This result is robust across all model, data, and dust choices.  Considering the numerous jet break models presented in \cite{Gao2013} (i.e., ``Phase 2'' to ``Phase 3''), we find only one that is consistent with our data.  This model requires slow-cooling electrons in a wind medium, and the electron energy spectral index, $p>2$ \citep[Tables 13/14 and 18/19 of][]{Gao2013}.  While these slope descriptions do not allow us to fully determine the order of the self-absorption frequency, $\nu_a$, cooling frequency, $\nu_{c}$, and characteristic frequency, $\nu_{m}$, (i.e., the pre- and post-break slopes are identical for $\nu_a < \min(\nu_m, \nu_c)$ and $\nu_m < \nu_a < \nu_c$ with these assumptions), in either case, $\nu_m < \nu_c$.  This model invokes $\alpha_1 = (3p-1)/4$, $\alpha_2 = (3p+1)/4$, and $\beta = (p-1)/2$ \citep{Gao2013}.  From our assumed spectral index, $\beta = 0.76 \pm 0.07$ \citep{Blanchard2024}, in this jet break interpretation, we would expect $\alpha_1 = (3\beta+1)/2 = 1.64\pm 0.10$ and $\alpha_2 = (3\beta + 2)/2 = 2.14 \pm 0.11$.  This provides an expected change in slope of $\Delta\alpha = 0.50 \pm 0.21$.  The expected parameters are in $\sim 1\sigma$ agreement with our measured $\alpha_1 = 1.51 \pm 0.03$, $\alpha_2 = 2.34 \pm 0.11$, and $\Delta\alpha = 0.83 \pm 0.14$. We therefore report $p = 2\beta+1 = 2.52 \pm 0.14$ for this decay model.  In the repeated analysis assuming the G23 model, we find a pre-break slope of $\alpha_1 = 1.45\pm 0.02$ and a post-break slope of $\alpha_2 = 2.45 \pm 0.10$, resulting in $\Delta \alpha = 1.00 \pm 0.12$.  These measured parameters are in $3\sigma$ consistency with the F99 measured parameters, and they are also in $\sim 1\sigma$ consistency with the predicted jet-break parameters listed prior. We find consistency with \cite{Laskar2023} who report $p = 2.53$, $\nu_m < \nu_c$, and predict a jet break at $t \sim 130$ days based on an observed 3 mm flux.  The interpretation in \cite{Levan2023} is incompatible with our interpretation, as they assume an early jet break at t $<$ 0.03 days and $p<2$.  In a recent radio-focused analysis of GRB\,221009A, \cite{Rhodes2024} did not find evidence for a jet break out to 475 days post-burst, however they do note that lateral structure in the jet could have prevented detection.  

Assuming equation (2) from \cite{Zhao2020} (and references therein), which describes the jet opening angle assuming a wind profile and using $E_{\gamma, \text{iso}} = 10^{55}$ erg \citep{Lesage23}; gamma-ray efficiency, $\eta = 0.2$; and wind parameter $A^* = 1$, a jet break at $\sim$ 50 days leads to a jet half-opening angle of $\theta_{jet}(\text{Wind}) = 4\degree$.  This is in contradiction to the earlier claims of a very narrow jet, \citep[e.g., $\sim 0.8 \degree, 1.5\degree$][respectively]{LHASSO23, Negro2023}.  If we instead use a wind parameter $A* = 10^{-3}$, we calculate $\theta_{jet}(\text{Wind}) = 0.7\degree$, which is in better agreement with early results.  This implies that the wind environment of GRB\,221009A has a much higher velocity or much lower mass-loss rate than the standardly assumed model.  Alternatively, \cite{OConnor23} suggests a two-component jet model, which would allow for two jet breaks. 

While we show support for the break at $t\sim 50$ days being interpreted as a jet break, there is also a break in the optical data at $\sim 1$ day \citep{Shrestha23}.  Using the ``Wind" assumptions (including $A^* = 1$) and equation above, a jet break at 1 day leads to a jet half-opening angle of $\theta_{jet}(\text{Wind}) = 1\degree$.  This break at $t\sim 1$ day, however, has also been interpreted as coming from the spectral ``cooling" break \citep{Laskar2023}.  Robust classification of  the $t\sim 50$ day optical break, especially in the context of earlier optical breaks, can only be achieved with modeling of the full multi-wavelength data set, which we leave for future work (Laskar et al., in prep.).

The jet half-opening angle and isotropic gamma-ray energy can be used to calculate the energy of the jet via $E_{\gamma} = E_{\gamma, iso} (1-cos(\theta_{jet}))$ \citep{Frail2001}.  We calculate $E_{\gamma} = 2.4 \times 10^{52}$ erg. \cite{Wang2018} found a mean $\log(E_{\gamma}/\text{erg}) = 49.54 \pm 1.29$ and a mean jet half-opening angle of $\theta_{jet} = 2.5 \pm 1\degree$ in an optical and X-ray study of 99 GRB afterglows.  Our measurements are in $\sim 1 \sigma$ consistency with the mean values.  This implies that while GRB\,221009A is extraordinary in observed properties, it is relatively average in intrinsic properties.

\subsubsection{Star Cluster or Dwarf Satellite Galaxy}

In the JWST NIRSpec and MIRI spectrum of the host galaxy at the location of the GRB, strong ionized hydrogen emission (H~\textsc{II}) and strong molecular hydrogen emission (H$_{2}$) were detected \citep{Blanchard2024}. H~\textsc{II} regions are ionized by the UV radiation from recently-formed massive stars.  Strong H~\textsc{II} emission is a sign of strong UV radiation, which could originate from many stars, i.e., a star cluster. Molecular hydrogen is also known to be associated with recent star formation \citep[e.g.,][]{Verma2024}.  The association of GRB\,221009A with a star cluster would be the first such detection, but it would not be in conflict with theoretical expectation from massive stellar collapse. GRB host galaxies at $z<1.2$ have been found to be bluer (i.e., UV-brighter) than core-collapse SN host galaxies, with the increased UV light thought to be caused by recent, intense star-formation \citep{Svensson2010}, and GRBs at $z<1.2$ are found to occur on the brightest parts of their host galaxy \citep{Fruchter2006, Svensson2010, Blanchard2016, Lyman2017}. In the context of host galaxy light distributions, the existence of a GRB within a star cluster (or otherwise dense, star-forming region) is, arguably, expected.

To model a young star cluster at the distance of GRB 221009A, we use the Flexible Stellar Population Synthesis \citep[FSPS;][]{ConroyGunnWhite2009, ConroyWhiteGunn2010, ConroyGunn2010} code with the assumptions of a gas-phase metallicity $Z = 0.12 Z_{\odot}$ and apparent magnitude $m_{F110W} = 24.84$ mag at z = 0.151 with $A_{V} = 4.6$ mag across the age range 1--4 Myr. This matches the gas-phase metallicity determined at the location of the GRB \citep{Blanchard2024}, and we normalize in the F110W band (as opposed to F160W or F814W), as it is our best measured quantity. We select this age range to force the star cluster to be sufficiently young to still have significant molecular hydrogen \citep{Hollyhead2015}. Under these assumptions, if the additional source is interpreted as a star cluster, it would have a mass of $M_{*} \sim 3 \times 10^5 M_{\odot}$. Star cluster masses have been well measured for clusters in the SMC, a similarly low metallicity and low SFR galaxy, and for this age range, there is not a single cluster with a mass $> 10^4 M_{\odot}$ \citep[see Figure 8 in][]{Hunter2003}.  A cluster of this mass and age in a low SFR \citep[0.17 $M_{\odot}/$yr,][]{Blanchard2024} galaxy is therefore unexpected.

We do not resolve the source in F814W, and we therefore calculate an upper limit on its radius of 140 pc.  A study of 164 star-forming dwarf galaxies at $z = 0.13 - 0.88$ found a range of effective radii $r_{e} = 0.1 - 6$ kpc (median $r_e = 1.2 \pm 2.3$ kpc) in the HST-ACS F814W bandpass \citep{Calabro2017}.  While a dwarf galaxy smaller than our limit of 140 pc is consistent with this distribution, it would have to be one of the smallest ever discovered.  Star clusters have linear diameters $\sim 20$ pc \citep{vandenBergh2006}, and are therefore consistent with our limit. Confirmation of this blue source as a star cluster, or otherwise dense star-forming region, can only come from continued blue flux detection after the afterglow and supernova have completely faded.

\subsubsection{Scattered-Light Echo}\label{subsubsec:lightecho}

A scattered light echo occurs when light from a supernova, or other bright transient, reflects off dust in the host galaxy of the transient.  The echo will have an optical light curve as a fading source with a change in decay rate at two times the light-travel time of the distance of the scattering dust. The change in slope is dependent on the distance of the scattering dust, the strength of the dust wind, and the presence of forward scattering \citep[][]{Chevalier1986}. The echo will appear in the spectrum with a SED shape nearly identical to the initially radiated light but with a $f_{\lambda}$ scaling law of $\lambda^{-\alpha}$, where $\alpha$ ranges from 1 - 2 \citep{Chevalier1986, Miller2010}.  It is expected that the SED of the echo is bluer than the transient.  Scattered light echoes have been detected for a number of SNe, including the well-studied and close core-collapse supernova, SN 1987A \citep{Crotts1995, Cikota2023}; a close and class-defining SNe Ia, SN 1991T \citep{Schmidt1994}; one of the most luminous core collapse supernovae, SN 2006gy \citep{Smith2008b, Miller2010}; and one of the closest normal Type Ia SNe, SN 2014J \citep{Crotts2015, Yang2018}.

A shallowing of the optical light curve from the afterglow decay, as we find tentative support for, is consistent with a scattered light echo.  We also see the SED becoming bluer with time.  Our HST and JWST data mostly probe the NIR, however the extinction corrected V3 F814W - F110W color is $-0.31 \pm 0.21$ mag.  This converts to a spectral slope, $\beta = 0.84 \pm 0.56$. The optical light at early times was dominated by the afterglow, with a spectral shape of $f_{\nu} \propto \nu^{-0.76 \pm 0.10}$ \citep{Blanchard2024}. The light echo spectral shape should therefore be $f_{\nu} \propto \nu^{1.24}$, assuming $f_{\lambda} \propto \lambda^{-2} \times f_{\lambda, AG}$, or $f_{\nu} \propto \nu^{0.24}$ assuming $f_{\lambda} \propto \lambda^{-1} \times f_{\lambda, AG}$.  These scalings produce expected F814W-F110W colors of $-0.45$ mag and $-0.08$ mag, respectively, in remarkable agreement with our measured color of $-0.31 \pm 0.21$ mag. 

\begin{figure}
\centering
\includegraphics[width = 0.49\textwidth]{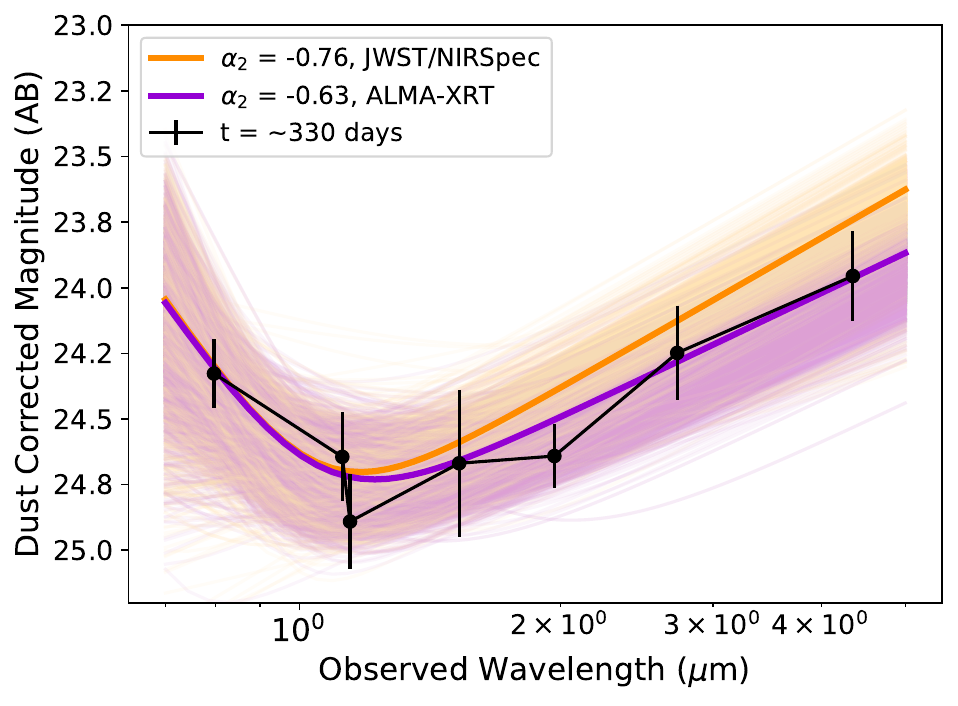}
\caption{A fit of a sum of two power laws to the $t \sim 330$ day SED.  In black is the HST and JWST data as presented in Table \ref{tbl:photometry}.  In purple and orange are the \texttt{emcee} best fits assuming the ALMA-XRT or JWST/NIRSpec afterglow slopes \citep{Blanchard2024} for the red power-law slope, respectively.  We also plot 3060 random pulls from the \texttt{emcee} posteriors to show the $1\sigma$ uncertainty in the fit.  The best fit blue slopes and their uncertainty are reported in Section \ref{subsubsec:lightecho}.\label{fig:PLSumSED}}
\end{figure}

We run an \texttt{emcee} fit of a sum of two power laws (similar to Equation (2), but without the SN component) to our dust-corrected $t \sim 330$ day data to further check the feasibility of this interpretation.  This fit is shown in Figure \ref{fig:PLSumSED}.  We assume a smoothness parameter, $s = -3$, to invert the break angle from the light curve modeling.  We run two fits: one with the redder power-law slope equal to the ALMA-XRT slope at $\Delta t = 190$ days (observer-frame) and the other with a red power-law slope equal to that measured from the JWST/NIRSpec spectrum at $\Delta t = 167$ days, both from \cite{Blanchard2024}.  We assume flat priors of 0 to 4.0 for the blue power-law slope and the two normalization scalings.  We measure blue slopes of $1.93^{+1.15}_{-0.86}$ and $1.96 ^{+1.10}_{-0.92}$, assuming the ALMA-XRT and JWST red slopes, respectively.  These slopes  convert to expected F814W-F110W colors of $-0.71 ^{+0.32}_{-0.43}$ mag and $-0.73^{+0.34}_{-0.41}$ mag, both in one sigma consistency with the expected light echo color range of $-0.08 - -0.45$ mag.  If this source continues to have an optical plateau while fading in the NIR bands, this could be confirmation of this blue source as a light echo.  If confirmed, this would mark the first detection of an optical light echo from a GRB afterglow.

\subsection{Comparison to Other GRBs}
GRB 221009A is exceptional in many regards.  In the interpretation that the break is a jet break,  we compare the jet break time of $t_{brk} \sim 50$ days to reported jet break times of all other GRBs (N = 138) through early-2020 reported in \cite{Zhao2020}.  130 of the 138 jet breaks in \cite{Zhao2020} are measured from X-ray or optical light curves, while the remaining eight were measured from radio light curves.  The median value of their sample of 138 jet break times is $\log(t_{brk, med}/s)  = 4.78 \sim 0.9$ days, and the standard deviation of their sample is $0.77$ dex. The jet break time of GRB\,221009A is later than any object in the sample, though is in $3\sigma$ agreement with the median.   With that said, the sample of \cite{Zhao2020} is observationally biased against GRBs with late ($t>12$ days) jet breaks, as most GRB afterglows are not observed in the X-ray or optical past $12$ days \citep{Melandri2014}. Indeed, only 5 of the 138 GRBs in the sample have jet break times later than 12 days.  The detection of a jet break at a superlatively late time in comparison to this sample is rather an observational consequence of the intrinsic brightness and closeness of the burst than an indication of the intrinsic rarity of or geometric uniqueness allowing for such a late break.

We also investigated the late-time ($t>80$ days) optical behavior of long GRBs.  We search for late-time ($t>80$ day) optical light curves for a subsample of long GRBs that are comparable to GRB\,221009A. We start with GRBs that are similarly highly energetic and include the 14 GRBs listed in \cite{Burns2023} with $E_{iso} > 2.5 \times 10^{54}$ erg (see their Table 4).  We add to this the three GRBs with detected VHE emission, GRBs 180720B \citep{Abdalla2019}, 190114C \citep{MAGIC2019a, MAGIC2019b}, and 190829A \citep{HESS2021}.  Finally we add to this the sample of four GRBs presented in \cite{Rastinejad24}: GRBs 030329, 100316D, 130427A, and 190829A, though we note that 190829A was already included in the VHE sample.  These GRBs were selected because they are close ($z<0.4$), have a detected SN, and importantly, have a well-populated $t>30$ days optical light curve.

In the case of the four presented in \cite{Rastinejad24}, all but GRB\,130427A are dominated by SN emission at $t = 4 - 100+$ days.  Following the fading of the SN associated with GRB\,130427A at $t\sim250$ days, the light curve appears to follow the decay pre-SN, excluding the possibility of a break in the interim.  GRB\,221009A is therefore unique among objects with late-time optical data in that its light curve does not appear to be dominated by supernova emission and has a $t\sim50$ day break.

In a literature search, we do not find optical data at $t>80$ days for the remainder of the GRBs in our comparison sample.  We note that for the afterglow of GRB\,160625B, \cite{Zhang2018} report $R$-band upper limits at $t \sim 58$ days (see their Figure 2), which was close to our ``late-time" cut off, and they find no evidence for a break.  We also note that for the afterglow of GRB\,110918A, \cite{Frederiks2013} present UVOT photometry to $\sim48$ days and found no evidence for a break.

\section{Conclusions and Future Work}

We present HST and JWST imaging of the afterglow and host galaxy of GRB\,221009A at $t>185$ days post-trigger.  We find, with high statistical significance, disagreement with the extension to late times of a simple power law afterglow plus supernova decay fit to early observations.  We next consider an alternative supernova plus afterglow decay model, which requires a break in the light curve.  This simple broken power-law + SN model under-predicts the latest F814W observations (Figure \ref{fig:3panel1}), which necessitates the addition of a secondary flux component in filters blueward of F160W.  In statistical AIC and BIC tests, we disfavor all models except the BPL + SN + C model, and we find these results are not sensitive to our choice of measurement for V1 F160W (see Section \ref{subsec:diffim}). In the model requiring the constant component, we find support for $t_{brk} \sim 50 \pm 10$ days which can be interpreted as a jet break assuming $\nu_m < \nu_c$ and $p>2$, consistent with the results presented in \cite{Laskar2023}.  The modeling also indicates SN\,2022xiw having an optical/NIR flux $ < 1.4 \times$ SN\,1998bw.  

In our analysis of the SED, we find evidence for a blue source in addition to the afterglow.  For this source to be explained exclusively by a SN similar to SN\,1998bw, it would need to be a factor of $\sim 3$ times more luminous than SN\,1998bw.  This is inconsistent with our upper scaling limit of $< 1.4$. We instead interpret this source as a scattered light echo.  We find consistency with this interpretation, with the predicted F814W-F110W color of $-0.45$ mag to $-0.08$ mag in agreement with our measured color of $-0.31 \pm 0.21$ mag.  If confirmed, this would be the first detection of an optical scattered light echo of a GRB afterglow.  In an alternative interpretation, the presence of strong molecular hydrogen emission at the location of the GRB \citep{Blanchard2024} lends support to the hypothesis that this source could be a young star cluster.  Under this assumption of $Z = 0.12Z_{\odot}$, $m_{F110W} = 24.84$ mag, and age $= 1-4$ Myr, we find a cluster with a mass of $M_{*} \sim 3\times 10^5 M_{\odot}$.  This mass is a factor of $\sim 10^2$ greater than masses of similarly young clusters in the SMC.  If confirmed, this would be one of the most massive young clusters in a low SFR environment. 

Assuming a continued power-law decay with $\alpha = 2.34$, the afterglow should fade below the HST and JWST detection limits within the next $\sim$year.  Continued monitoring of this source in the filters presented here will provide the observations necessary to better constrain the nature of the additional source.  In later time light curves, a light echo is expected to appear as a fading blue source, while a star cluster or dwarf satellite galaxy is expected to appear as a constant source in all wavelengths (though brightest in the bluest filters).

Template imaging will be crucial in robustly separating the host galaxy and diffraction spike light from the transient light. While observations in the HST and JWST bands will be helpful in better analyzing the early optical and NIR data, to fully understand GRB\,221009A, a full multi-wavelength analysis with all epochs of data will be necessary.

\section{Acknowledgments}

We thank the anonymous referee for comments which improved this manuscript.  We thank Wen-fai Fong, Charlie Kilpatrick, and Jay Strader for valuable conversations regarding this manuscript.  We thank Jenni Barnes, Yvette Cendes, Dan Kasen, Joel Leja, Dan Siegel, and S. Karthik Yadavalli for significant contributions to the proposals which provided these data.  This research made use of Photutils, an Astropy package for detection and photometry of astronomical sources \citep{larry_bradley_2023_7946442}.  T.E. is supported by NASA through the NASA Hubble Fellowship grant HST-HF2-51504.001-A awarded by the Space Telescope Science Institute, which is operated by the Association of Universities for Research in Astronomy, Inc., for NASA, under contract NAS5-26555. W.J-G is supported by NASA through the NASA Hubble Fellowship grant HSTHF2-51558.001-A awarded by the Space Telescope Science Institute, which is operated by the Association of Universities for Research in Astronomy, Inc., for NASA, under contract NAS5-26555. B.D.M. acknowledges support from NASA through the Astrophysics Theory Program (grant AST-2406637) and the Simons Investigator Program (grant 727700).  The Flatiron Institute is supported by the Simons Foundation. Support for program GO-17278 was provided by NASA through a grant from the Space Telescope Science Institute, which is operated by the Association of Universities for Research in Astronomy, Inc., under NASA contract NAS 5–26555.  Support for program DD-2784 was provided by NASA through a grant from the Space Telescope Science Institute, which is operated by the Association of Universities for Research in Astronomy, Inc., under NASA contract NAS 5-03127. The data were obtained from the Mikulski Archive for Space Telescopes at the Space Telescope Science Institute, which is operated by the Association of Universities for Research in Astronomy, Inc., under NASA contract NAS 5-03127 for JWST. This research has made use of the NASA/IPAC Extragalactic Database (NED), which is funded by the National Aeronautics and Space Administration and operated by the California Institute of Technology.

\vspace{5mm}
\facilities{HST (WFC3), JWST (NIRCam)}
\software{Astrodrizzle \citep{AstroDrizzle}, Astropy \citep{astropy:2013, astropy:2018, astropy:2022}, corner \citep{corner}, GALFIT \citep{Peng2010}, HOTPANTS \citep{Becker2015}, matplotlib \citep{matplotlibref}, numpy \citep{numpyref}, photutils \citep{larry_bradley_2023_7946442}, \texttt{Source Extractor} \citep{SourceExtractor}}

\bibliography{myrefs}
\bibliographystyle{aasjournal}

\end{document}